\documentclass[format=sigconf]{acmart}
\usepackage{graphicx}
\usepackage{epstopdf}
\usepackage{epsfig}
\usepackage{color}
\usepackage{url}
\usepackage{pseudocode}
\usepackage{algorithm}
\usepackage{algpseudocode}
\usepackage{listings}
\usepackage{etoolbox}
\usepackage{subcaption}
\usepackage{ragged2e}
\usepackage{mathtools}
\usepackage{caption}
\usepackage{stmaryrd}
\usepackage{amsmath}
\usepackage{pifont}
\usepackage{enumitem}
\usepackage{amssymb}
\usepackage{hyperref}
\usepackage{mdframed}
\usepackage{xspace}			
\hypersetup{
colorlinks=true, 
linktoc=page,     
linkcolor=[rgb]{0.0,0.0,0.8},  
urlcolor=[rgb]{0,0.0,0.8},
citecolor=[rgb]{0,0.0,1.0}
}

\expandafter\def\expandafter\UrlBreaks\expandafter{\UrlBreaks%
  \do\a\do\b\do\c\do\d\do\e\do\f\do\g\do\h\do\i\do\j%
  \do\k\do\l\do\m\do\n\do\o\do\p\do\q\do\r\do\s\do\t%
  \do\u\do\v\do\w\do\x\do\y\do\z\do\A\do\B\do\C\do\D%
  \do\E\do\F\do\G\do\H\do\I\do\J\do\K\do\L\do\M\do\N%
  \do\O\do\P\do\Q\do\R\do\S\do\T\do\U\do\V\do\W\do\X%
\do\Y\do\Z}

\usepackage{tikz}

\lstdefinestyle{Cstyle}{
  language=C,                     
  numbersep=5pt,                  
  backgroundcolor=\color{white},  
  showspaces=false,               
  showstringspaces=false,         
  showtabs=false,                 
  tabsize=2,                      
  captionpos=b,                   
  breaklines=true,                
  breakatwhitespace=true,         
  title=\lstname,                 
  basicstyle=\footnotesize\ttfamily,
  numbers=left,                   
  xleftmargin=5.0ex,              
  keywordstyle=\color{magenta},
  commentstyle=\color[rgb]{0,0.6,0},
  stringstyle=\color{codepurple},
  numberstyle=\tiny\ttfamily\color{gray},
  rulecolor=\color{gray}
}
\lstdefinestyle{Clargestyle}{
  language=C,                     
  numbersep=5pt,                  
  backgroundcolor=\color{white},  
  showspaces=false,               
  showstringspaces=false,         
  showtabs=false,                 
  tabsize=2,                      
  captionpos=b,                   
  breaklines=true,                
  breakatwhitespace=true,         
  title=\lstname,                 
  basicstyle=\scriptsize\ttfamily,
  numbers=left,                   
  xleftmargin=5.0ex,              
  keywordstyle=\color{magenta},
  commentstyle=\color[rgb]{0,0.6,0},
  stringstyle=\color{codepurple},
  numberstyle=\small\ttfamily\color{gray},
  rulecolor=\color{gray}
}

\lstdefinestyle{plain}{
  language=C,                     
  numbersep=5pt,                  
  backgroundcolor=\color{white},  
  showspaces=false,               
  showstringspaces=false,         
  showtabs=false,                 
  tabsize=2,                      
  captionpos=b,                   
  breaklines=true,                
  breakatwhitespace=true,         
  title=\lstname,                 
  basicstyle=\small\ttfamily,
  numbers=none,                   
  xleftmargin=5.0ex,              
  keywordstyle=\color{magenta},
  commentstyle=\color[rgb]{0,0.6,0},
  stringstyle=\color{codepurple},
  numberstyle=\tiny\ttfamily\color{gray},
  rulecolor=\color{gray}
}

\def\eg{{e.g.,}\xspace}
\def\ie{{i.e.,}\xspace}
\def\etc{\emph{etc.}\xspace}
\def\etal{{et al.}\xspace}

\newif \ifcomments
\commentstrue

\ifcomments
\newcommand{\jason}[1]{{--\textcolor{red}{#1}--}}
\newcommand{\suman}[1]{{---\textcolor{blue}{#1}---}}
\newcommand{\theo}[1]{{---\textcolor{brown}{#1}---}}
\else
\newcommand{\jason}[1]{}
\newcommand{\suman}[1]{}
\newcommand{\theo}[1]{}
\fi

\def\bzip{\texttt{bzip2}\xspace}

\def\sortingsize{64 bytes\xspace}

\def\pname{\textsc{SlowFuzz}\xspace}

\begin{document}
\copyrightyear{2017}
\acmYear{2017}
\setcopyright{licensedusgovmixed}
\acmConference{CCS '17}{October 30-November 3, 2017}{Dallas, TX, USA}\acmPrice{15.00}\acmDOI{10.1145/3133956.3134073}
\acmISBN{978-1-4503-4946-8/17/10}

\title{\pname: Automated Domain-Independent Detection of Algorithmic Complexity Vulnerabilities}

\author{Theofilos Petsios}
\email{theofilos@cs.columbia.edu}
\affiliation{Columbia University}
\author{Jason Zhao}
\email{zhao.s.jason@columbia.edu}
\affiliation{Columbia University}
\author{Angelos D. Keromytis}
\email{angelos@cs.columbia.edu}
\affiliation{Columbia University}
\author{Suman Jana}
\email{suman@cs.columbia.edu}
\affiliation{Columbia University}

\maketitle
\subsection*{Abstract}
Algorithmic complexity vulnerabilities occur when the worst-case time/space
complexity of an application is significantly higher than the respective
average case for particular user-controlled inputs. When such conditions are
met, an attacker can launch Denial-of-Service attacks against a
vulnerable application by providing inputs that trigger the
worst-case behavior.  Such attacks have been known to have serious effects on
production systems, take down entire websites, or lead to bypasses of Web
Application Firewalls.

Unfortunately, existing detection mechanisms for algorithmic complexity
vulnerabilities are domain-specific and often require significant manual
effort.  In this paper, we design, implement, and evaluate \pname, a
domain-independent framework for automatically finding algorithmic complexity
vulnerabilities. \pname automatically
finds inputs that trigger worst-case algorithmic behavior in the tested binary.
\pname uses resource-usage-guided evolutionary search techniques to
automatically find inputs that maximize computational resource utilization for a
given application. 

We demonstrate that \pname successfully generates inputs that match the
theoretical worst-case performance for several well-known algorithms. \pname was also able
to generate a large number of inputs that trigger different algorithmic
complexity vulnerabilities in real-world applications, including various zip
parsers used in antivirus software, regular expression libraries used in Web
Application Firewalls, as well as hash table implementations used in Web
applications. In particular, \pname generated inputs that achieve 300-times
slowdown in the decompression routine of the \bzip utility, discovered regular
expressions that exhibit matching times exponential in the input size, and also
managed to automatically produce inputs that trigger a high number of collisions
in PHP's default hashtable implementation.



\section{Introduction}
\label{sec:intro}
Algorithmic complexity vulnerabilities result from large differences between the
worst-case and average-case time/space complexities of algorithms
or data structures used by affected software~\cite{dos_via_complexity}. An
attacker can exploit such vulnerabilities by providing specially crafted inputs
that trigger the worst-case behavior in the victim software to launch
Denial-of-Service (DoS) attacks. For example, regular expression matching is
known to exhibit widely varying levels of time complexity (from linear to
exponential) on input string size depending on the type of the regular
expression and underlying implementation details. Similarly, the run times of
hash table insertion and lookup operations can differ
significantly if the hashtable implementation suffers from a large
number of hash collisions. Sorting algorithms like quicksort can have
an $O(nlogn)$ average-case complexity but an $O(n^2)$ worst-case complexity.
Such worst-case behaviors have been known to take down entire
websites~\cite{StackExc3:online}, disable/bypass Web Application
Firewalls (WAF)~\cite{cve-2011-5021}, or to keep thousands of CPUs busy by merely
performing hash-table insertions~\cite{phpdos,WhydoesS96:online}.

Despite their potential severity, in practice, detecting algorithmic complexity
vulnerabilities in a domain-independent way is a hard, multi-faceted problem. It is often infeasible
to completely abandon algorithms or data structures with high worst-case
complexities without severely restricting the functionality or
backwards-compatibility of an application.  Manual time complexity analysis of
real-world applications is hard to scale. Moreover, asymptotic complexity
analysis ignores the constant factors that can significantly affect the
application execution time despite not impacting the overall complexity class. All
these factors significantly harden the detection of algorithmic complexity
vulnerabilities.

Even when real-world applications use well-understood algorithms, time
complexity analysis is still non-trivial for the following reasons. First, the
time/space complexity analysis changes significantly even with minor
implementation variations (for instance, the choice of the pivot in the
quicksort algorithm drastically affects its worst-case runtime
behavior~\cite{clrs}). Reasoning about the effects of such changes requires
significant manual effort.  Second, most real-world applications often have
multiple inter-connected components that interact in complex ways. This
interconnection further complicates the estimation of the overall complexity,
even when the time complexity of the individual components is well understood.

Most existing detection mechanisms for algorithmic complexity vulnerabilities
use domain- and implementation-specific heuristics or rules, \eg detect
excessive backtracking during regular expression matching~\cite{backtracking,
berglund2014analyzing}. However, such rules tend to be brittle and are hard to
scale to a large number of diverse domains, since their creation and maintenance
requires significant manual effort and expertise. Moreover, keeping such rules
up-to-date with newer software versions is onerous, as even minor changes to the
implementation might require significant changes in the rules.

In this work, we design, implement, and evaluate a novel dynamic
domain-independent approach for automatically finding inputs that trigger
worst-case algorithmic complexity vulnerabilities in tested applications.
In particular, we introduce \pname, an evolutionary-search-based framework that
can automatically find inputs to maximize resource utilization (instruction count,
memory usage \etc) for a given test binary. \pname is fully automated and
does not require any manual guidance or domain-specific rules. The key idea
behind \pname is that the problem of finding algorithmic complexity
vulnerabilities can be posed as an optimization problem whose goal is to
find an input that maximizes resource utilization of a target application. We
develop an evolutionary search technique specifically designed to find
solutions for this optimization problem.


We evaluate \pname on a variety of real world applications, including the PCRE
library for regular expression matching~\cite{pcre}, the \bzip
compression/decompression utility, as well as the hash table implementation
of PHP. We demonstrate that \pname can successfully generate inputs that
trigger complexity vulnerabilities in all the above contexts. Particularly, we
show that \pname generates inputs that achieve a 300-times slowdown when
decompressed by the \bzip utility, can produce regular expressions that exhibit
matching times exponential in the input's size, and also manages to
automatically generate  inputs that trigger a high number of collisions in real-world
PHP applications. We also demonstrate that our
evolutionary guidance scheme achieves more than 100\% improvement over
code coverage at steering input generation towards triggering complexity vulnerabilities.


In summary, this work makes the following contributions:
\begin{itemize}[leftmargin=*]
    \item We present \pname, the first, to the best of our knowledge,
        domain-independent dynamic testing tool for automatically finding
        algorithmic complexity vulnerabilities without any manual guidance.
    \item We design an evolutionary guidance engine with novel mutation schemes particularly fitted
        towards generating inputs that trigger worst-case resource usage
        behaviors in a given application. Our scheme achieves more than 100\%
        improvement over code-coverage-guided input generation at finding such inputs.
    \item We evaluate \pname on a variety of complex real-world applications
        and demonstrate its efficacy at detecting complexity vulnerabilities
        in diverse domains including large real-world software like the \bzip utility and the PCRE
        regular expression library.
\end{itemize}

The rest of the paper is organized as follows. We provide a high-level overview of \pname's inner workings with a motivating
example in Section~\ref{sec:motivateeg}. We describe the details of our methodology in Section~\ref{sec:method}.
The implementation of \pname is described in Section~\ref{sec:impl} and the evaluation results are presented
in Section~\ref{sec:eval}. Section~\ref{sec:discussion} outlines the limitations of our current prototype and discusses possible future extensions.
Finally, we discuss related work in Section~\ref{sec:related} and conclude in
Section~\ref{sec:concl}.

\section{Overview}
\label{sec:motivateeg}

\subsection{Problem Description}
\label{subsec:threat_model}
In this paper, we detect algorithmic complexity vulnerabilities in a given
application by detecting inputs that cause large variations in resource
utilization through the number of executed instructions or CPU usage
for all inputs of a given size. We assume that our tool has gray-box access to
the application binary, i.e., it can instrument the binary in order to harvest
different fine-grained resource usage information from multiple runs of the
binary, with different inputs. Note that our goal is not to estimate the
asymptotic complexities of the underlying algorithms or data structures of the
application. Instead, we measure the resource usage variation in some pre-defined
metric like the total edges accessed during a run, and try to maximize that metric.
Even though, in most cases,
the inputs causing worst-case behaviors under such metrics will be the ones
demonstrating the actual worst-case \textit{asymptotic} behaviors, but this
may not always be true
due to the constant factors ignored in the asymptotic time complexity, the small
input sizes, \etc

\noindent
{\bf Threat model.}
Our threat model assumes that an attacker can provide arbitrary
specially-crafted inputs to the vulnerable software to trigger worst-case
behaviors. This is a very realistic threat-model as most non-trivial real-world
software like Web applications and regular expression matchers need to deal with
inputs from untrusted sources. For a subset of our experiments involving
regular expression matching, we assume that attackers can control regular
expressions provided to the matchers. This is a valid assumption for a large
set of applications that provide search functionality through custom regular
expressions from untrusted users.

\subsection{A Motivating Example}
\label{subsec:example}

In order to understand how our technique works, let us consider quicksort, one
of the simplest yet most widely used sorting algorithms.  It is
well-known~\cite{clrs} that quicksort has an average time complexity of
$O(nlogn)$ but a worst-case complexity of $O(n^2)$ where $n$ is the size of the
input.  However, finding an actual input that demonstrates the worst-case
behavior in a particular quicksort implementation depends on low-level details
like the pivot selection mechanism. If an adversary knows the actual pivot
selection scheme used by the implementation, she can use domain-specific rules
to find an input that will trigger the worst-case behavior (\eg the quadratic
time complexity)~\cite{quicksortkiller}.

However, in our setting, \pname does not know any domain-specific rules. It
also does not understand the semantics of pivot selection or which part of the
code implements the pivot selection logic, even though it has access to the
quicksort implementation. We would still like \pname to generate inputs that
trigger the corresponding worst-case behavior and identify the algorithmic
complexity vulnerability.

This brings us to the following research question: \textit{how can \pname
automatically generate inputs that would trigger worst-case performance in a
tested binary in a domain-independent manner?} The search space of all inputs
is too large to search exhaustively. Our key intuition in this paper is that
evolutionary search techniques can be used to iteratively find inputs that are
closer to triggering the worst-case behavior. Adopting an evolutionary testing
approach, \pname begins with a corpus of seed inputs, applies mutations to each
of the inputs in the corpus, and ranks each of the inputs based on their
resource usage patterns.  \pname keeps the highest ranked inputs for further
mutations in upcoming generations.

\begin{figure}
\lstset{basicstyle=\footnotesize, style=Clargestyle, gobble=-6}
\begin{minipage}{0.95\textwidth}
    \begin{lstlisting}[label=lst:quicksort]
function quicksort(array):
    /* initialize three arrays to hold
    elements smaller, equal and greater
    than the pivot */
    smaller, equal, greater = [], [], []
    if len(array) <= 1:
        return
    pivot = array[0]
    for x in array:
        if x > pivot:
            greater.append(x)
        else if x == pivot:
            equal.append(x)
        else if x < pivot:
            smaller.append(x)
    quicksort(greater)
    quicksort(smaller)
    array = concat(smaller, equal, greater)
\end{lstlisting}
\vspace{-4pt}
\end{minipage}
\vspace{-4pt}
\centering
    \includegraphics[width=0.5\columnwidth]{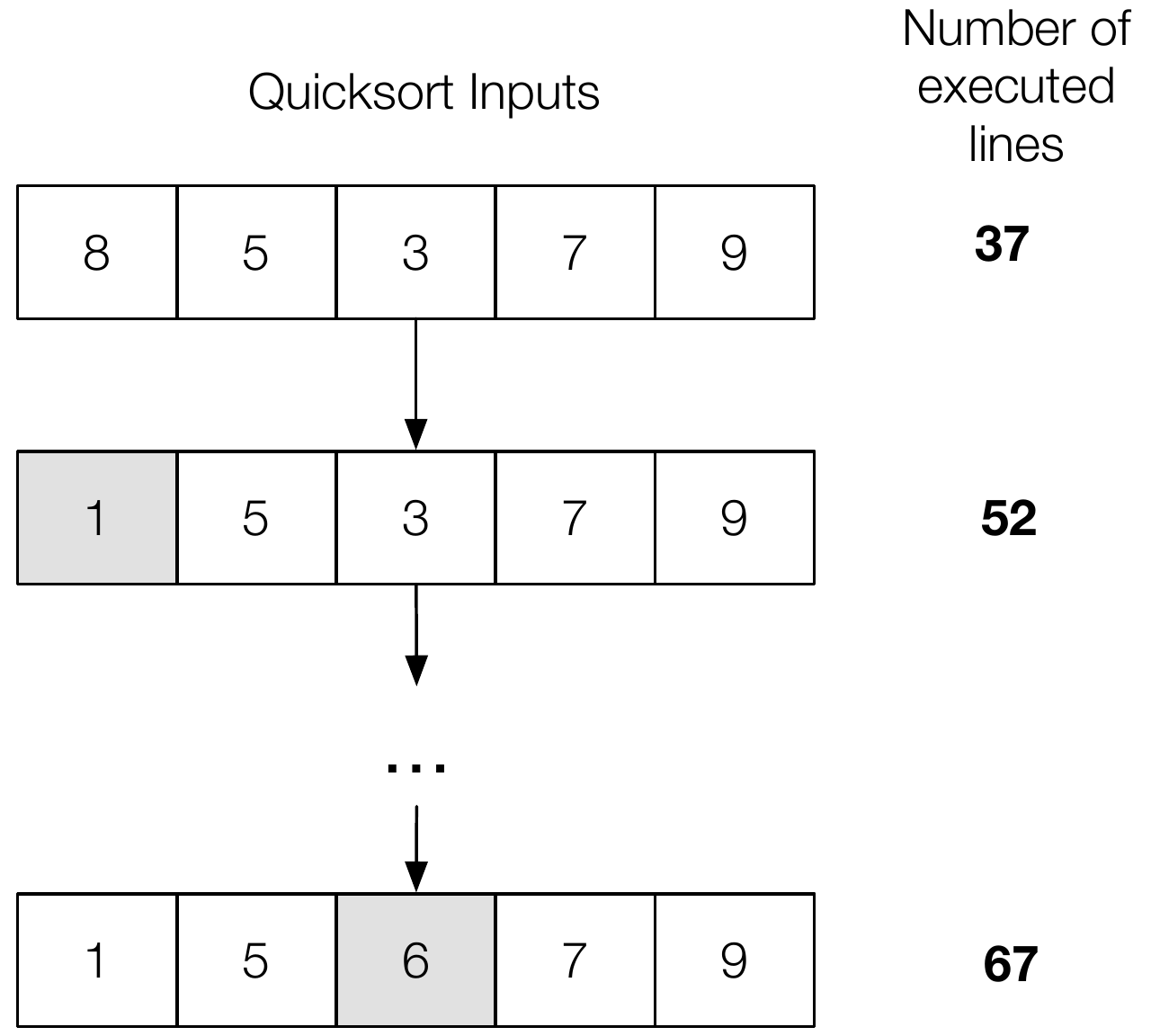}

\caption{Pseudocode for quicksort with a simple pivot selection mechanism and overview of \pname's evolutionary
search process for finding inputs that demonstrate worst-case quadratic time complexity. The shaded boxes indicate
mutated inputs.}
\label{fig:quicksort}
\end{figure}

To further illustrate this point, let us consider the pseudocode of
Figure~\ref{fig:quicksort}, depicting a quicksort example with a simple pivot
selection scheme\textemdash the first element of the array being selected as the pivot.
In this case, the worst-case behavior can be elicited by an already sorted array.
Let us also assume that \pname's initial corpus consists of some arrays of numbers
and that none of them are completely sorted. Executing this quicksort implementation with
the seed arrays will result in a different number of statements/instructions executed based
on how close each of these arrays are to being sorted. \pname will assign a score to each of these inputs
based on the number of statements executed by the quicksort implementation for
each of the
inputs. The inputs resulting in the highest number of executed statements will be selected for
further mutation to create the next generation of inputs. Therefore, each upcoming generation
will have inputs that are closer to being completely sorted than the inputs of the previous generations.

For example, let us assume the initial corpus for \pname consists of a single array
$\mathcal{I}=[8, 5, 3, 7,9]$. At each step, \pname picks at random
an input from the corpus, mutates it, and passes the mutated input to the
above quicksort implementation while recording the number of executed statements.
As shown in Figure~\ref{fig:quicksort}, the input [8, 5, 3, 7, 9] results in the execution
of $37$ lines of code (LOC). Let us assume that this input is mutated into
$[1, 5, 3, 7, 9]$ that causes the execution of $52$ LOC which is higher than
the original input and therefore [1, 5, 3, 7, 9] is selected for further mutation.
Eventually, \pname will find a completely sorted array (e.g., [1, 5, 6, 7, 9] as shown
in Figure~\ref{fig:quicksort}) that will demonstrate the worst-case quadratic
behavior. We provide a more thorough analysis of \pname's performance on various
sorting implementations in Section~\ref{subsec:microbench}.

\section{Methodology}
\label{sec:method}

The key observation for our methodology is that evolutionary search techniques
together with dynamic analysis present a promising approach for finding
inputs that demonstrate worst-case complexity of a test application in a domain-independent
way. However, to enable \pname to efficiently find such inputs, we need to
carefully design effective guidance mechanisms and mutation schemes to drive
\pname's input generation process. We design a new evolutionary algorithm with
customized guidance mechanisms and mutation schemes that are tailored for
finding inputs causing worst-case behavior.


Algorithm~\ref{alg:slowtest} shows the core evolutionary engine of \pname.
Initially, \pname randomly selects an input to execute from a given seed corpus
(line 4), which is mutated (line 5) and passed as input to the test application
(line 6).  During an execution, profiling info such as the different types of
resource usage of the application are recorded (lines 6-8). An input is
scored based on its resource usage and is added to the mutation corpus
if the input is deemed as a slow unit (lines 9-12).

\begin{algorithm}[ht!]
\algrenewcommand{\alglinenumber}[1]{\color{gray}\footnotesize#1:}
    \caption{SlowFuzz: Report all slow units for application
    $\mathcal{A}$ after $n$ generations, starting from a corpus
    $\mathcal{I}$}
\begin{algorithmic}[1]

    \Procedure{DiffTest}{$\mathcal{I}$, $\mathcal{A}$, $n$, $GlobalState$}
    \State $units=\emptyset$  ;reported slowunits
    \While{$generation \leq n$ and $\mathcal{I} \neq \emptyset$}
        \State $input=$ \textsc{RandomChoice}($\mathcal{I}$)
        \State $mut\_input=$ \textsc{Mutate}($input$)
        \State $app\_insn, app\_outputs =~$\textsc{Run}($\mathcal{A}, mut\_input$)
        \State $gen\_insn~\cup= \{app\_insn\}$
        \State $gen\_usage~\cup= \{app\_usage\}$
        \If{$\textsc{SlowUnit}(gen\_insn, gen\_usage$, \par
        \hskip 6.7em $GlobalState)$}
            \State $\mathcal{I} \leftarrow \mathcal{I} \cup mut\_input$
            \State $units ~\cup= mut\_input$
        \EndIf
        \State $generation=generation + 1$
    \EndWhile
    \State \Return $units$
    \EndProcedure
\end{algorithmic}
\label{alg:slowtest}
\end{algorithm}

In the following Sections, we describe the core components of \pname's engine, particularly the fitness function used to determine
whether an input is a slow unit or not, and the offset and type of mutations performed on each of the individual
inputs in the corpus.

\subsection{Fitness Functions}
\label{subsec:guidance}
As shown in Algorithm~\ref{alg:slowtest}, \pname determines, after each
execution, whether the executed unit should be considered for further mutations
(lines 9-12). \pname ranks the current inputs based on the scores assigned to them by
a fitness function and keeps the fittest ones for further mutation. Popular coverage-based
fitness functions which are often used by evolutionary fuzzers to detect crashes,
are not well suited for our purpose as they do not consider loop iterations
which are crucial for detecting worst-case time complexity.

\pname's input generation is guided by a fitness function based on resource
usage. Such a fitness function is generic and can take into consideration different kinds
of resource usage like CPU usage, energy, memory, etc. In order to measure the CPU
usage in a fine-grained way, \pname's fitness function keeps track of the total count of
all instructions executed during a run of a test program. The intuition is that
the test program becomes slower as the number of executed instructions increases.
Therefore, the fitness function selects the inputs that result in the highest number of
executed instructions as the slowest units. For efficiency, we monitor execution
at the basic-block level instead of instructions while counting the total number of executed
instructions for a program. We found that this method is more effective
at guiding input generation than directly using the time taken by the test program to run.
The runtime of a program shows large variations, depending on the application's
concurrency characteristics or other programs that are executing in the same CPU,
and therefore is not a reliable indicator for small increases in CPU
usage.

\subsection{Mutation Strategy}
\label{subsec:mutations}

\pname introduces several new mutation strategies tailored to identify inputs
that demonstrate the worst-case complexity of a program. A mutation strategy decides
which mutation operations to apply and which byte offsets in an input to modify,
to generate a new mutated input (Algorithm~\ref{alg:slowtest}, line 5).

\pname supports the following mutation operations: (i) add/remove a new/existing byte from the input; ii) randomly modify a
bit/byte in the input; iii) randomly change the order of a subset of the input
bytes; iv) randomly change bytes whose values are within the range
of ASCII codes for digits (\ie 0x30-0x39); v) perform a crossover operation in
a given buffer mixing different parts of the input; and vi) mutate bytes solely
using characters or strings from a user-provided dictionary.

We describe the different mutation strategies supported by \pname below.
Section~\ref{subsubsec:mut_engines} presents a detailed performance comparison
of these strategies.

\noindent
{\bf Random Mutations.}
\label{subsec:mut_random}
Random mutations are the simplest mutation strategy supported by \pname.
Under this mutation strategy, one of the aforementioned
mutations is selected at random and is applied on an input, as long as it does not violate
other constraints for the given testing session, such as exceeding the maximum
input length specified by the auditor. This strategy is similar to the ones used by popular
evolutionary fuzzers like AFL~\cite{afl} and libFuzzer~\cite{libFuzzer} for finding crashes
or memory safety issues.

\vspace{2pt}
\noindent
{\bf Mutation priority.}
Under this strategy, the mutation operation is selected with $\epsilon$ probability based on its success at producing slow units
during previous executions. The mutation operation is picked at random with $(1-\epsilon)$ probability. In contrast, the mutation
offset is still selected at random just like the strategy described above.

In particular, during testing, we count all the cases in which a mutation operation resulted in an increase in the
observed instruction count and the number of times that operation has been selected. Based on these values,
we assign a score to each mutation operation denoting the probability of the mutation to be successful at increasing
the instruction count. For example, a score of $0$ denotes that the mutation operation has never resulted in an
increase in the number of executed instructions, whereas a score of $1$ denotes that the mutation always
resulted in an increase.

We pick the highest-scoring mutation among all mutation operations with a probability $\epsilon$.
The tunable parameter $\epsilon$ determines how often a mutation  operation will be selected at random
versus based on its score. Essentially, different values of $\epsilon$ provide different trade-offs between
exploration and exploitation. In \pname, we set the default value of
$\epsilon$ to $0.5$.

\vspace{2pt}
\noindent
{\bf Offset priority.}
\label{subsec:mut_offset}
This strategy selects the mutation operation to be applied randomly
at each step, but the offset to be mutated is selected based on prior
history of success at increasing the number of executed instructions.
The mutation offset is selected based on the results of previous executions
with a probability $\epsilon$ and at random with a probability $(1-\epsilon)$.
In the first case, we select the offset that showed the most promise based on
previous executions (each offset is given a score ranging from 0 to 1 denoting the percentage
of times in which the mutation of that offset led to an increase in the number of instructions).

\vspace{2pt}
\noindent
{\bf Hybrid.}
\label{subsec:mut_hybrid}
In this last mode of operation we apply a combination of both mutation and
offset priority as described above. For each offset, we maintain an array of
probabilities of success for each of the mutation operations that are being performed.
Instead of maintaining a coarse-grained success probability for each mutation
in the mutation priority strategy, we maintain fine-grained success probabilities for each
offset/mutation operation pairs. We compute the score
of each offset by computing the average of success probabilities of all mutation operations
at that offset. During each mutation, with a probability of $\epsilon$, we pick the offset and operation
with the highest scores. The mutation offset and operation are also picked randomly with a probability
of $(1-\epsilon)$.

\section{Implementation}
\label{sec:impl}

The \pname prototype is built on top of libFuzzer~\cite{libFuzzer}, a popular
evolutionary fuzzer for finding crash and memory safety bugs. We outline the implementation
details of different components of \pname below. Overall, our
modifications to libFuzzer consist of 550 lines of C++ code. We used
Clang v4.0 for compiling our modifications along with the rest of libFuzzer code.

\begin{figure}[t] \centering
\includegraphics[width=\columnwidth]{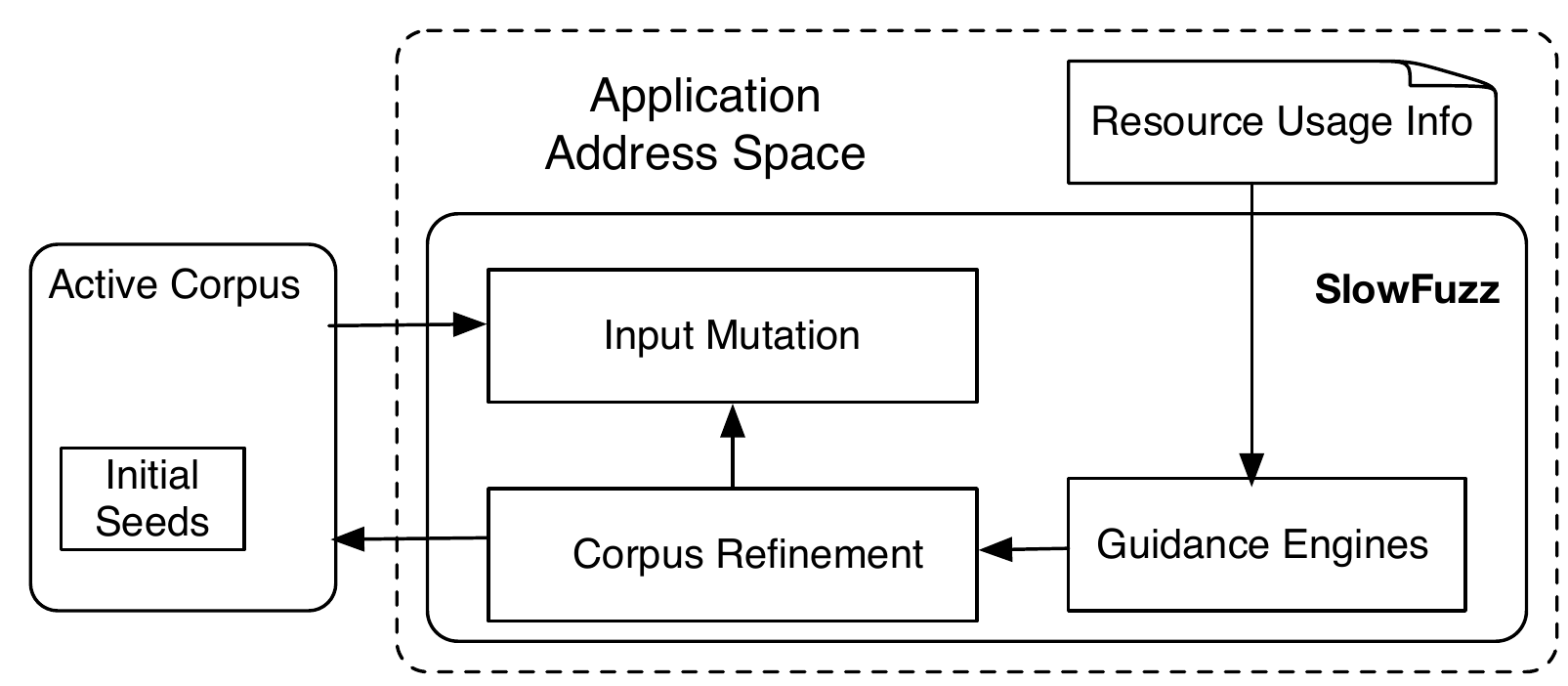}
    \caption{\pname architecture.}
    \label{fig:architecture}
\end{figure}

Figure~\ref{fig:architecture} shows \pname's high-level architecture. Similar to
the popular evolutionary fuzzers like AFL~\cite{afl} and
libFuzzer~\cite{libFuzzer}, \pname executes in the same address space as the
application being tested.  We instrument the test application so that \pname can
have access to different resource usage metrics (e.g, number of instructions executed)
needed for its analysis. The instrumented test application
subsequently is executed under the control of \pname's analysis engine.  \pname
maintains an active corpus of inputs to be passed into the tested applications and
refines the corpus during execution based on \pname's fitness function.
For each generation, an input is selected, mutated, then passed into the
\texttt{main} routine of the application for its execution.

\noindent
{\bf Instrumentation.}
\label{subsubsec:instrumentation}
Similar to libFuzzer, \pname's instrumentation is based on Clang's
SanitizerCoverage~\cite{sanitizercoverage} passes. Particularly,
SanitizerCoverage allows tracking of each executed function, basic block, and
edge in the Control Flow Graph (CFG). It also allows us to register callbacks
for each of these events. \pname makes use of SanitizerCoverage's eight bit counter
capability that maps each Control Flow Graph (CFG) edge into an eight bit counter
representing the number of times that edge was accessed during an
execution. We use the counter to keep track of the following ranges: 1, 2, 3, 4-7, 8-15, 16-31,
32-127, 128+. This provides a balance between accuracy of the counts and the overhead
incurred for maintaining them. This information is then passed into \pname's fitness function,
which determines whether an input is slow enough to keep for the next generation of mutations.

\noindent
{\bf Mutations.}
\label{subsubsec:mutations}
LibFuzzer provides API support for  custom input mutations. However, in order to
implement the mutation strategies proposed in Section~\ref{subsec:mutations}, we
had to modify libFuzzer internals. Particularly, we
augment the functions used in libFuzzer's \texttt{Mutator} class to
return information on the mutation operation, offset, and the range of affected bytes for
each new input generated by LibFuzzer. This information is used to compute
the scores necessary for supporting mutation piority, offset priority, and hybrid
modes as described in Section~\ref{subsec:mutations} without any additional runtime overhead.

\section{Evaluation}
\label{sec:eval}

In this Section, we evaluate \pname on the following objectives:
a) Is \pname capable of generating inputs that match the theoretical worst-case
complexity for a given algorithm's implementation? b) Is \pname capable of efficiently
finding inputs that cause performance slowdowns in real-world applications? c) How do
the different mutation and guidance engines of \pname affect its performance? d) How
does \pname compare with code-coverage-guided search at finding inputs demonstrating
worst-case application behavior?

We describe the detailed results of our evaluation  in the following Sections. All our
experiments were performed on a machine with 23GB of RAM, equipped with an
Intel(R) Xeon(R) CPU X5550 @ 2.67GHz and running 64-bit Debian 8 (jessie),
compiled with GCC version 4.9.2, with a kernel version 4.5.0. All binaries were
compiled using the Clang-4.0 compiler toolchain. All instruction counts and
execution times are measured using the Linux \texttt{perf} profiler v3.16,
averaging over 10 repetitions for each \texttt{perf} execution.

\subsection{Overview}
\label{subsec:overview}

In order to adequately address the questions outlined in the previous
Section, we execute \pname on applications of different algorithmic profiles
and evaluate its ability of generating inputs that demonstrate worst case behavior.

First, we examine if \pname generates inputs that demonstrate the
theoretical worst-case behavior of well-known algorithms. We apply \pname
on sorting algorithms with well-known complexities. The results are presented in
Section~\ref{subsec:microbench}. Subsequently, we
apply \pname on different applications and algorithms that have been known
to be vulnerable to complexity attacks: the PCRE regular
expression library, the default hash table implementation of PHP, and the
\bzip binary. In all cases, we demonstrate that \pname is able to trigger complexity
vulnerabilities. Table~\ref{table:result_summary} shows a summary of our findings.

\begin{table}[ht!]
\centering
\begin{tabular}{l|l}
    \textbf{Tested Application} & \textbf{Fuzzing Outcome} \\
\hline
Insertion sort~\cite{clrs} & 41.59x slowdown \\
Quicksort (Fig~\ref{fig:quicksort}) & 5.12x slowdown \\
Apple quicksort  & 3.34x slowdown \\
OpenBSD quicksort & 3.30x slowdown \\
NetBSD quicksort  & 8.7\% slowdown \\
GNU quicksort  & 26.36\% slowdown\\
    PCRE (fixed input) & 78 exponential \&  \\
    & 765 superlinear regexes\\
PCRE (fixed regex) & 8\% - 25\% slowdown \\
PHP hashtable & 20 collisions in 64 keys\\
\bzip decompression & \textasciitilde 300x slowdown \\
\end{tabular}
\caption{Result Summary}
\label{table:result_summary}
\end{table}

As shown in Table~\ref{table:result_summary}, \pname is successful at inducing
significant slowdown on all tested applications. Moreover, when applied to the PCRE library, it managed to
generate regular expressions that exhibit exponential and super-linear (worse
than quadratic) matching automatically, without any knowledge of the structure of
a regular expression. Likewise, it successfully generated inputs that induce
a high number of collisions when inserted into a PHP hash table, without any
notion of hash functions. In the following Sections, we provide details
on each of the above test settings.

\subsection{Sorting}
\label{subsec:microbench}

\begin{figure}[ht!]
\includegraphics[width=\columnwidth]{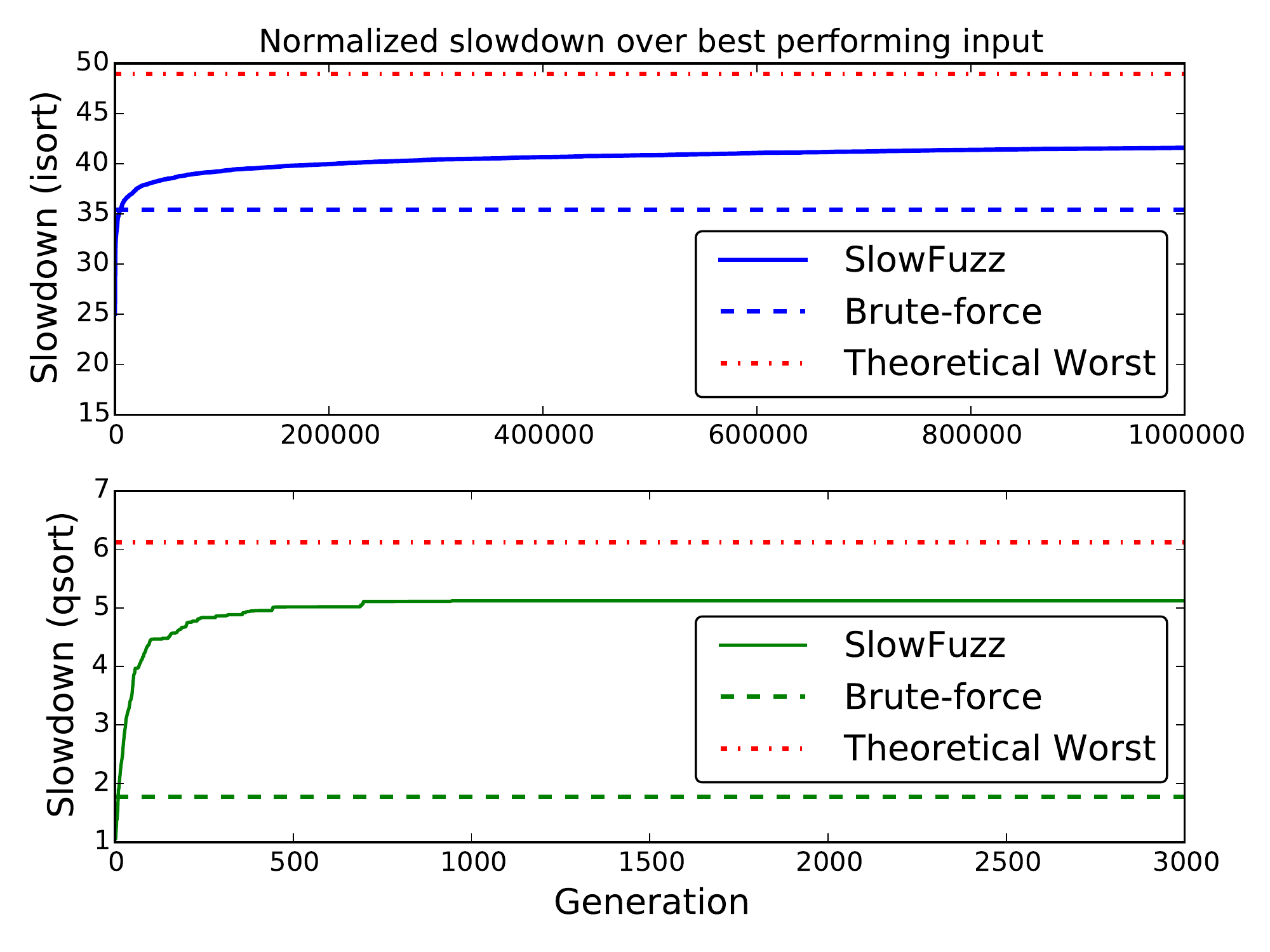}
    \caption{Best slowdown achieved by \pname at each generation (normalized
    over the slowdown of the best-performing input) versus best random
    testing outcome, on our insertion sort and quicksort drivers,
    for an input size of 64 bytes (average of 100 runs). The \pname achieves slowdowns of
84.97\% and 83.74\% compared to the theoretical worst cases for insertion sort and quicksort respectively.}
\label{fig:isort_ref}
\end{figure}

\noindent
{\bf Simple quicksort and insertion sort.}
Our first evaluation of \pname's consistency with theoretical results is
performed on common sorting algorithms with well-known worst-performing inputs.
To this end, we initially apply \pname on an implementation of the
insertion sort algorithm~\cite{clrs}, as well as on an implementation of
quicksort~\cite{clrs} in which the first sub-array element is always selected as
the pivot. Both of the above implementations demonstrate quadratic complexity
when the input passed to them is sorted.
We run \pname for 1 million generations on the
above implementations, sorting a file with a size of \sortingsize, and examine the
slowdown \pname introduced over the fastest unit seen during testing. To do so,
we count the total instructions executed by each program for each of the
inputs, subtracting all instructions not relevant to the quicksort
functionality (\eg loader code). Our results are presented in Figure~\ref{fig:isort_ref}.


%

Figure~\ref{fig:isort_ref} represents an average of 100 runs. In each run,
\pname started execution with a single random 64 byte seed, and executed for
1 million generations.
We notice that \pname achieves 41.59x and 5.12x slowdowns for insertion sort
and quicksort respectively. In order to examine how this behavior compares
to random testing, we randomly generated 1 million inputs of 64 bytes each and
measured the instructions required for insertion sort and quicksort,
respectively.  Figure~\ref{fig:isort_ref} depicts the \textit{maximum} slowdown
achieved through random testing \textit{across all} runs. We notice that in
both cases \pname outperforms the brute-force worst-input estimation. Finally,
we observe that the gap between brute-force search and \pname is much higher
for quicksort than insertion, which is consistent with the fact
that average case complexity of insertion sort is $O(n^2)$, compared to
quicksort's $O(nlogn)$. Therefore, a random input is more likely to demonstrate
worst-case behavior for insertion sort but not for quicksort.

\noindent
{\bf Real-world quicksort implementations.}
We also examined how \pname performs when applied
to real-world quicksort implementations. Particularly, we applied it to the
Apple~\cite{appleqsort}, GNU~\cite{gnuqsort}, NetBSD~\cite{netbsdqsort},
and OpenBSD~\cite{openbsdqsort} quicksort implementations. We notice
that \pname's performance on real world implementations is consistent with the
quicksort performance that we observed in the experiments described above.
In particular, the slowdowns generated by \pname were (in increasing order)
8.7\%, for theNetBSD implementation, 26.36\% for the GNU quicksort implementation,
3.30x for the OpenBSD implementation and 3.34x for the Apple implementation. We
notice that, despite the fact these implementations use
efficient pivot selection strategies, \pname still manages to trigger
significant slowdowns. On the contrary, repeating the same experiment using
naive coverage-based fuzzing yields slowdowns that never surpass 5\% for any of
the libraries. This is an expected result, as coverage-based fuzzers are geared
towards maximizing coverage, and thus do not favor inputs exercising the same edges
repeatedly over inputs that discover new edges.

\begin{figure}[ht!]
\includegraphics[width=\columnwidth]{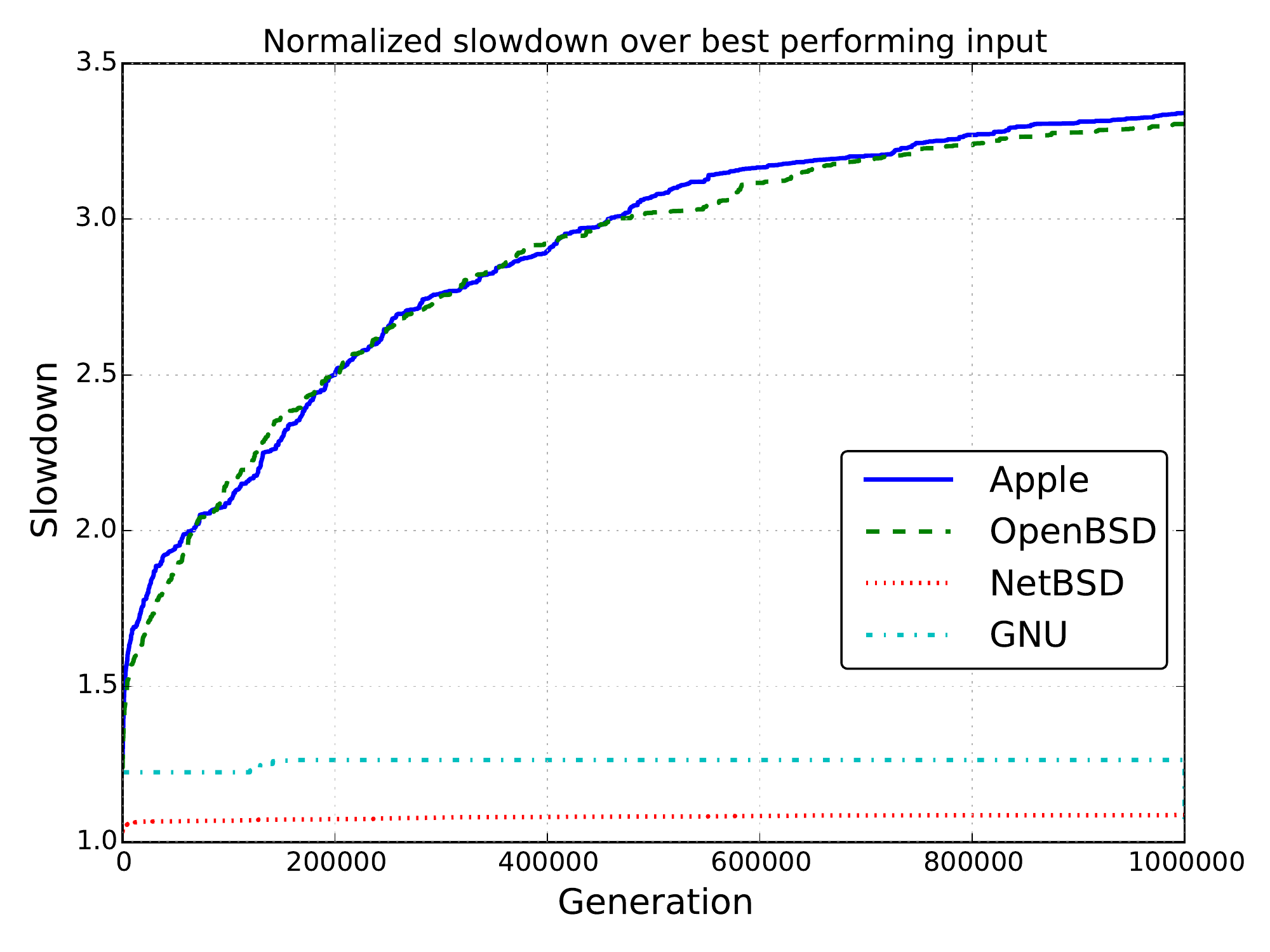}
    \caption{Best slowdown (with respect to the best-performing input) achieved
    by \pname at each generation normalized over the best random testing outcome, on
    real-world quicksort implementations, for an input size of 64 bytes
    (average of 100 runs).}
\label{fig:qsort_oses}
\end{figure}

Finally, we note that, similar to the experiment of
Figure~\ref{fig:isort_ref}, the slowdowns for Figure~\ref{fig:qsort_oses} are
also measured in terms of executed instructions, normalized over the instructions of the best
performing input seen during testing.



\vspace{0.1in}
\begin{mdframed}
    \textbf{Result 1}: \pname was able to generate inputs for quicksort and
    insertion sort that achieve 83.74\% and 84.97\% of the theoretical
    worst-case, respectively without any information on the algorithm internals.
\end{mdframed}
\vspace{0.1in}

\subsection{Regular Expressions}
\label{subsec:micro_regex}
Regular expression implementations are known to be
susceptible to complexity attacks~\cite{WhydoesS96:online,Regulare13:online,cve-2013-4287}.
In particular, there are over $150$ Regular expression Denial of Service (ReDoS)
vulnerabilities registered in the National Vulnerability Database (NVD), which
are the result of exponential (\eg ~\cite{cve-2015-2526}) or
super-linear (worse than quadratic) \eg ~\cite{cve-2013-2099} complexity of regular expression matching
by several existing matchers~\cite{rexploiter}.

Even performing domain-specific analyses of whether an application is
susceptible to ReDoS attacks is non-trivial. Several works are solely dedicated
to the detection of exploitation of such vulnerabilities.
Recently, Rexploiter~\cite{rexploiter}
presented algorithms to detect whether a given regular
expression may result in non-deterministic finite automata (NFA) that require
super-linear or exponential matching times for specially crafted inputs. They have
also presented domain-specific algorithms to generate inputs
capable of triggering such worst-case performance. The above denote the
hardness of \pname's task, namely finding regular expressions
that may result in super-linear or exponential matching times without any domain
knowledge.

\begin{figure}[ht!]
\includegraphics[width=\columnwidth]{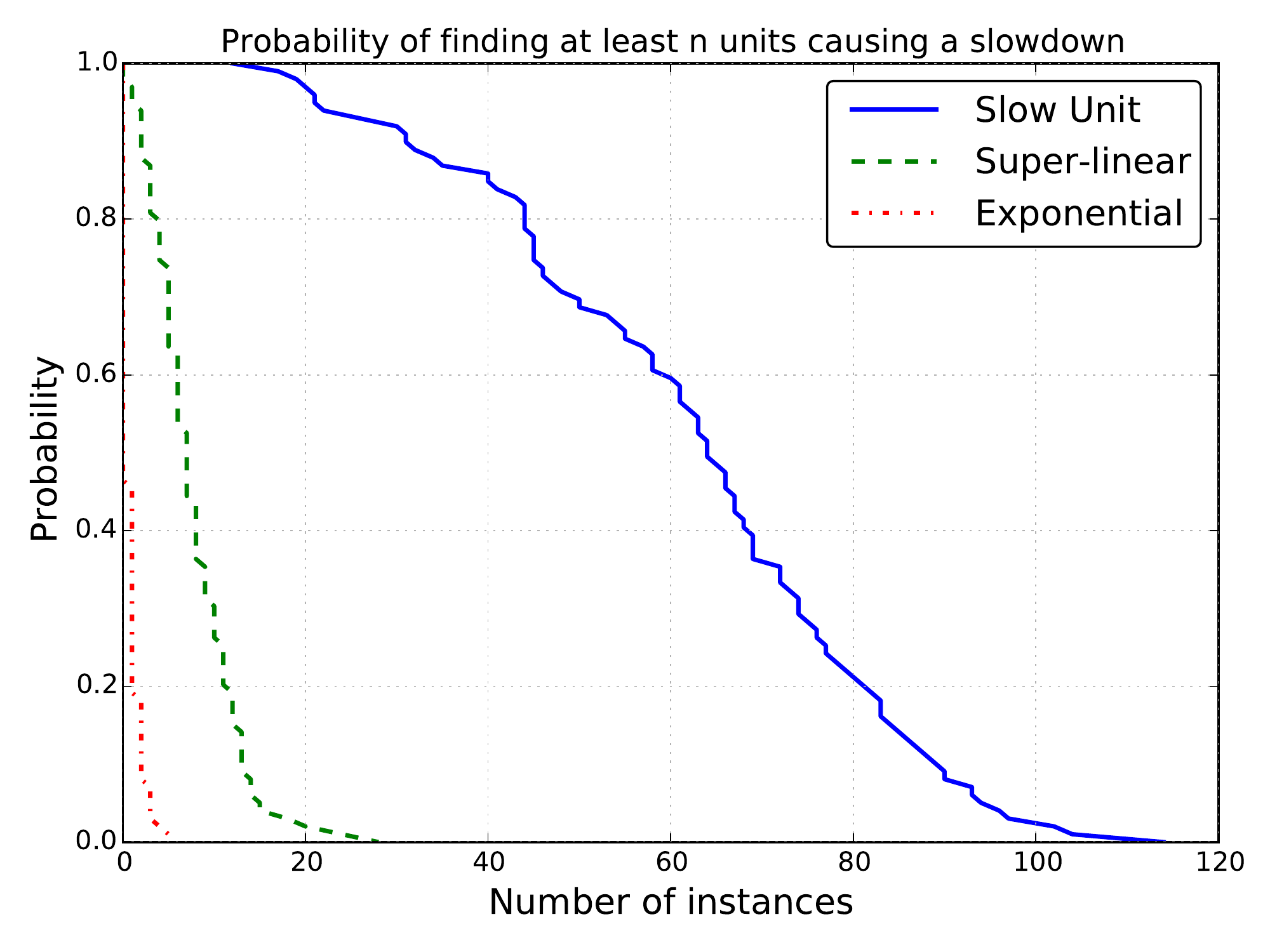}
    \caption{Probability of \pname
    finding at least $n$ unique instances of regexes that cause a slowdown, or
    exhibit super-linear and exponential matching times, after 1 million
    generations (inverse CDF over 100 runs).}
    \label{fig:regex_ecdf}
\end{figure}

For the regular expression setting we perform two separate experiments to check whether \pname can produce
i) regular expressions which
exhibit super-linear and exponential matching times,
ii) inputs that cause slowdown during matching, given a fixed regular expression.
To this end, we apply \pname on
the PCRE regular expression library~\cite{pcre} and provide it with a
character set of the symbols used in PCRE-compliant regular expressions (in the
form of a dictionary). Notice that we do not further guide \pname with
respect to what mutations should be done and \pname's engine is completely
agnostic of the structure of a valid regular expression. In all cases, we
start testing from an empty corpus without providing any seeds of regular
expressions to \pname.

\noindent
{\bf Fixed string and mutated regular expressions.} For the first part of our evaluation, we apply \pname on a binary that
utilizes the PCRE library to perform regular expression matching and
we let \pname mutate the regular expression part of the
\texttt{pcre2\_match} call used for the matching, using a dictionary
of regular expression characters.
The input to be matched against the regular expression is selected from a
random pool of characters and \pname executes for a total of 1 million
generations, or until a time-out is hit. The regular expressions
generated by \pname are kept limited to 10 characters or less. Once a \pname
session ends, we evaluate the time complexity of the generated regular expressions
utilizing~\texttt{Rexploiter}~\cite{rexploiter}, which detects if the regular
expression is super-linear, exponential, or none of the two. We repeat
the above process for a total of 100 fuzzing sessions.

Overall, \pname generates a total of 33343 regular expressions during the above
100 sessions, out of which 27142 are rejected as invalid whereas 6201 are valid
regular expressions that caused a slowdown. Out of the valid regular expressions, 765 are
superlinear and 78 are exponential. This experiment demonstrates that despite being
agnostic of the semantics of regex matching, \pname successfully generates regexes
requiring super-linear and exponential matching times. Six such examples are presented in
Table~\ref{table:regex_example}.

\begin{table}[ht!]
\centering
\begin{tabular}{c|c}
Super-linear (greater than quadratic) & Exponential \\
\hline
c*ca*b*a*b & (b+)+c \\
a+b+b+b+a+ & c*(b+b)+c \\
c*c+ccbc+  & a(a|a*)+a\\
\end{tabular}
\caption{Sample regexes generated by \pname resulting in
    super-linear (greater than quadratic) and exponential matching complexity.}
\label{table:regex_example}
\end{table}

\begin{figure}[ht!]
\includegraphics[width=0.55\columnwidth]{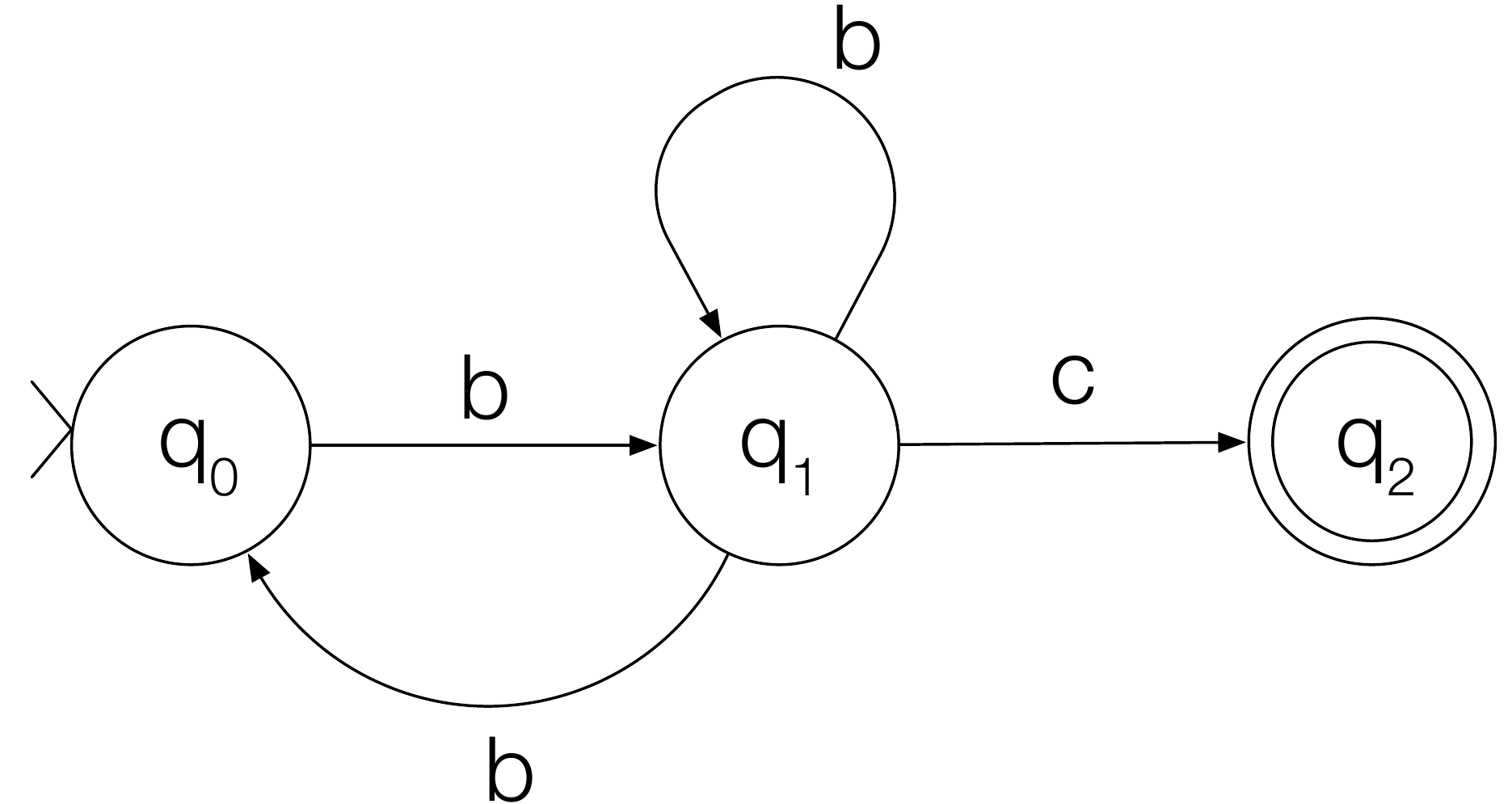}
    \caption{NFA for the regular expression (b+)+c suffering from
    exponential matching complexity as found by \pname. $q_0$ is the entry
    state, $q_2$ the accept state, and $q_1$ the pivot state
    for the exponential backtracking.}
\label{fig:nfa}
\end{figure}

\noindent
{\bf A detailed case study.} The regexes presented in Table~\ref{table:regex_example} are typical examples
of regular expressions that require non-linear matching running times. This happens
due to the existence of different paths in the respective NFAs, which reach the same state
through an identical sequence of labels. Such paths have a devastating effect during backtracking~\cite{rexploiter}.
To further elaborate on this property, let us consider the NFA depicted in Figure~\ref{fig:nfa}, which
corresponds to the regular expression (b+)+c of Table~\ref{table:regex_example}.

We notice that, for the NFA shown in Figure~\ref{fig:nfa}, starting from state $q_1$, it is possible
to reach $q_1$ again, through two different paths, namely the paths
$(q_1 \xrightarrow{\text{b}} q_0, q_0 \xrightarrow{\text{b}} q_1)$ and
$(q_1 \xrightarrow{\text{b}} q_1, q_1 \xrightarrow{\text{b}} q_1)$.
Moreover, we notice that the labels in the transitions for both of the above paths
are the same: 'bb' is consumed in both cases. Thus, as it is possible to reach $q_2$ from
$q_1$ (via label c) as well as reach $q_1$ from the initial state $q_0$, there will be an exponentially
large number of paths to consider in the case of backtracking. Similar issues arise with loops appearing in
NFAs with super-linear matching~\cite{rexploiter}.

As mentioned above, on average, among the valid regular expressions
generated by \pname, approximately 12.33\% of the regexes have
super-linear matching complexity, whereas 2.29\% on average have exponential
matching complexity. The aforementioned results are aggregates across all the
100 executions of the experiment. In order to estimate the probability
 of \pname to generate a regex that exhibits a
slowdown~\footnote{Notice that due to \pname's guidance engine, any regex
produced must exhibit increased instruction count as compared to \textit{all} previous
regexes.}, or super-linear and exponential matching times in a
\textit{single} session, we calculate the respective inverse CDF
which is shown in Figure~\ref{fig:regex_ecdf}.
We notice that, for all the regular expressions observed, \pname successfully
generates inputs that incur a slowdown during matching. In particular, with
90\% probability, \pname generates at least 2 regular expressions requiring
super-linear matching time and at least 31 regular expressions that cause a slowdown.
\pname generates at least one regex requiring exponential matching time with a
probability of 45.45\% .

\begin{figure}[ht!]
\includegraphics[width=\columnwidth]{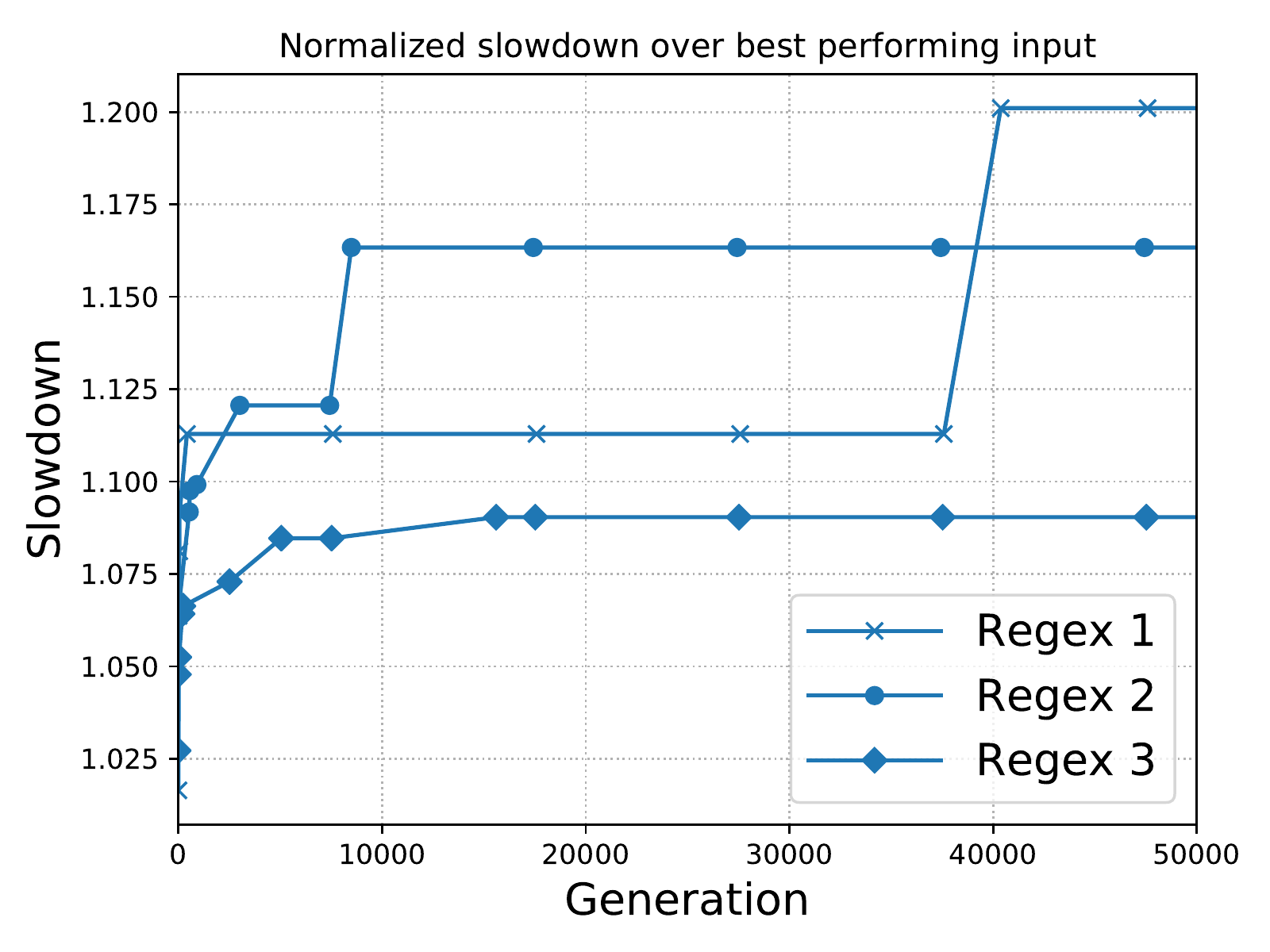}
    \caption{Best slowdown achieved by \pname-generated input strings (normalized
    over the slowdown of the best-performing input), when
    matching against fixed regular expressions used in WAFs (normalized against
    best performing input over an average of 100 runs). The corresponding regexes are listed
in Appendix~\ref{sec:waf_regexes}.}
\label{fig:all_wafs}
\end{figure}

\noindent
{\bf Fixed regular expression and mutated string.} In the second part of our
evaluation of \pname on regular expressions, we
seek to examine if, for a given \textit{fixed} regular expression, \pname is
able to generate inputs that can introduce a slowdown during matching. We collect
PCRE-compliant regular expressions from popular Web Application
Firewalls (WAF)~\cite{attacker90:online}, and utilized the PCRE library to
match input strings generated by \pname against each regular
expression. For this experiment, we apply \pname on a total of 25
regular expressions, and we record the total instructions executed by the PCRE
library when matching the regular expression against \pname's generated units, at
each generation. For our set of regular expressions, \pname achieved monotonically
increasing slowdowns, ranging from 8\% to 25\%. Figure~\ref{fig:all_wafs} presents
how the slowdown varies as fuzzing progresses, for three representative regex
samples with different slowdown patterns.

\subsection{Hash Tables}
\label{subsec:micro_hash}
Hash tables are a core data structure in a wide variety of software. The performance of
hash table lookup and insertion operations significantly affects the overall application performance.
Complexity attacks against hash table implementations may induce unwanted effects
ranging from performance slowdowns to full-fledged DoS
~\cite{phpdos,WhydoesS96:online,Regulare13:online,cve-2013-4287,cve-2015-2526}.
In order to evaluate if \pname can
generate inputs that trigger collisions without knowing any details about the underlying hash functions,
we apply it on the hash table implementation of PHP (v5.6.26), which is known to be vulnerable to collision
attacks.

\noindent
{\bf PHP Hashtables.}
Hashtables are prevalent in PHP and they also serve as the backbone for
PHP's array interface. PHP v5.x utilizes the DJBX33A hash function for hashing using
string keys, which can bee seen in Listing~\ref{lst:djbx33a}.

We notice that for two strings of the form `ab' and `cd' to collide, the
following property must hold~\cite{phpcollisions}:
\begin{center}
$c = a + n \wedge d = b - 33 * n, n\in \mathbb{Z}$
\end{center}
It is also easy to show that if two equal-length strings A and B collide, then
strings xAy, xBy where x and y are any prefix and suffix respectively, also collide.
Using the above property, one can construct a worst-case performing sequence of
inputs~\cite{phphashdos}, forcing a worst-case insertion time of $O(n^2)$.

\lstset{basicstyle=\scriptsize, style=Cstyle, gobble=-6}
\begin{lstlisting}[caption={DJBX33A hash without loop unrolling.},
                   label=lst:djbx33a]
/*
 * @arKey is the array key to be hashed
 * @nKeyLenth is the length of arKey
 */
static inline ulong
zend_inline_hash_func(const char *arKey, uint nKeyLength)
{
		register ulong hash = 5381;

		for (uint i = 0; i < nKeyLength; ++i) {
				hash = ((hash << 5) + hash) + arKey[i];
		}

		return hash;
}
\end{lstlisting}
Abusing the complexity characteristics of the BJBX33A hash, attackers performed
DoS attacks against PHP, Python and Ruby applications in 2011. As a response,
PHP added an option in its \texttt{ini} configuration
to set a limit on the number of collisions that are allowed to happen. However,
in 2015, similar DoS attacks~\cite{dos2015} were reported, abusing PHP's JSON
parsing into hash tables. In this experiment we examine how \pname performs when
applied to this particular hash function implementation.

\begin{figure}[ht!]
\includegraphics[width=\columnwidth]{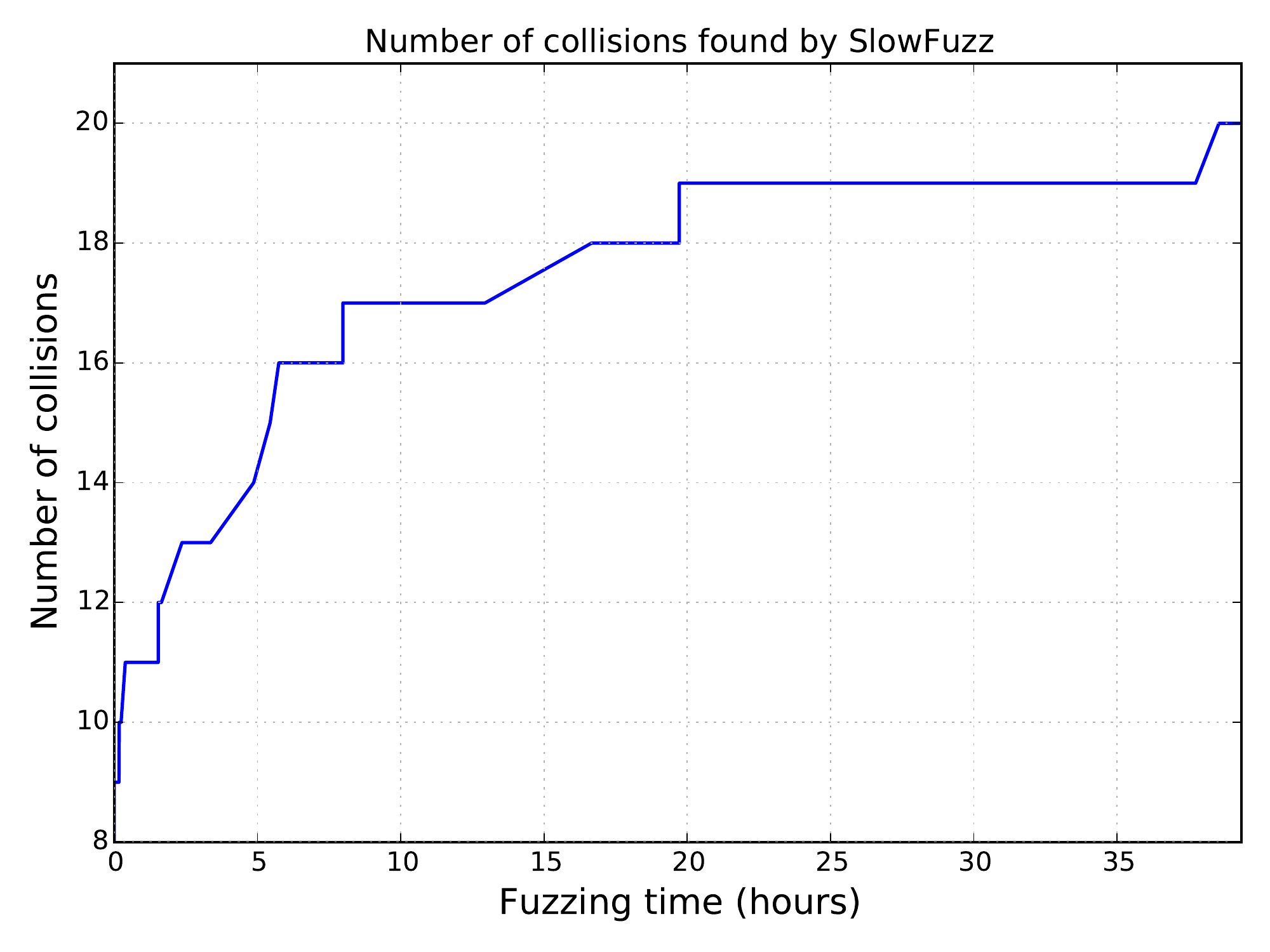}
    \caption{Number of collisions found by \pname per generation, when applying
    it on the PHP 5.6 hashtable impelemntation, for at most of 64 insertions
    with string keys.}
\label{fig:php_hash}
\end{figure}

Our experimental setup is as follows: we ported the PHP hash table
implementation so that the latter can be used in any C/C++ implementation,
removing all the interpreter-specific variables and macros, however leaving all
the non interpreter-related components intact. Subsequently, we created a
hash table with a size of 64 entries, and utilized \pname to perform
\textit{a maximum} of 64 insertions to the hash table, using strings as keys,
starting from a corpus consisting of a single input that causes 8 collisions.
In particular, the keys for the hash table insertions were provided by \pname
at each generation and \pname evolved its corpus of strings using a hybrid mutation
strategy. Given a hash table of 64 entries and 64 insertions to the hash table,
the maximum number of collisions that can be performed is also 64. In order
to measure the number of collisions occurring in the hashtable at each generation,
we created a PHP module (running in the context of PHP), and measured the number
of collisions induced by each input that \pname generates. We perform our measurements
after the respective elements are inserted into a \textit{real} PHP array. Our
results are presented in Figure~\ref{fig:php_hash}.

We notice that despite the complex conditions required to trigger a hash
collision and without knowing any details about the hash function, \pname's
evolutionary engine reaches 31.25\% of the theoretical worst-case after
approximately 40 hours of fuzzing, using a single CPU.
\pname's stateful, evolutionary
guidance achieves monotonically increasing slowdowns, despite the complex
constraints imposed by the hash function.  On the contrary, repeating the same
experiment using coverage-based fuzzing, yielded non-monotonically increasing
collisions, and at no point an input was generated causing more than 8
collisions.  In particular, fuzzing using coverage generated 58 inputs with a
median of 5 collisions.

\subsection{ZIP Utilities}
\label{subsec:macro_zip}

Zip utilities that support various compression/decompression schemes are another
instance of applications that have been shown to suffer from Denial of Service
attacks. For instance, an algorithmic complexity vulnerability used in the
sorting algorithms in the \bzip application~\footnote{The vulnerability is
found in \texttt{BZip2CompressorOutputStream} for Apache Commons Compress
before 1.4.1} allowed remote attackers to cause DoS via increased CPU
consumption, when they provided a file with many repeating
inputs~\cite{cve-2012-2098}.

\begin{figure}[ht!]
\includegraphics[width=\columnwidth]{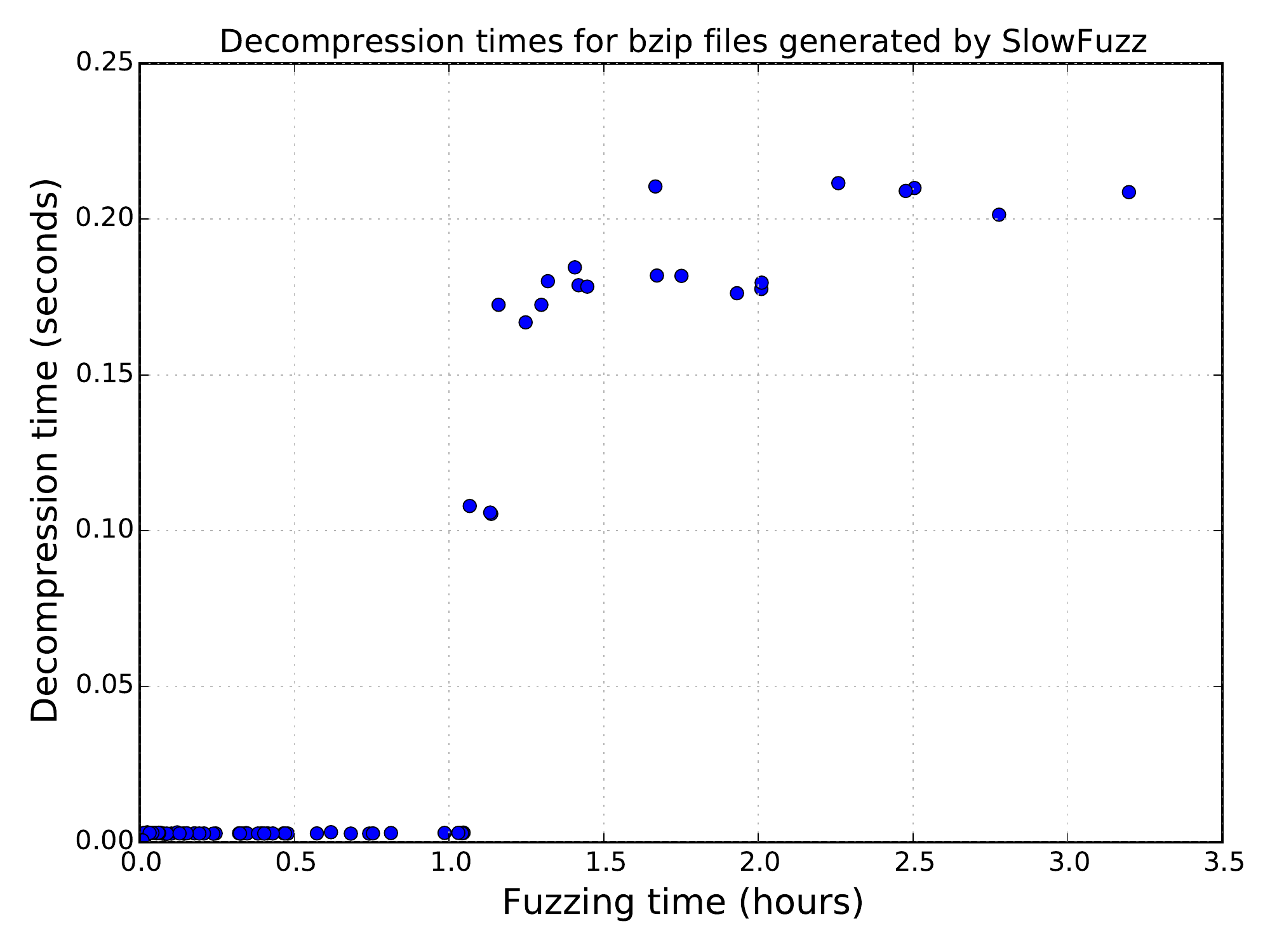}
    \caption{Slowdowns observed while decompressing inputs generated by \pname
    using the \bzip binary. The maximum file size is set to 250 bytes.}
    \label{fig:bzip}
\end{figure}

In order to evaluate how \pname performs when applied to the
compression/decompression libraries, we apply it on \bzip v1.0.6. In
particular, we utilize \pname to create compressed files of a maximum of 250
bytes, and we subsequently use the libbzip2 library to decompress them. Based
on the slowdowns observed during decompression, \pname evolves its input
corpus, mutating each input using its hybrid mode of operation. Our
experimental results are presented in Figure~\ref{fig:bzip}.

\noindent
{\bf A detailed case study.} Figure~\ref{fig:bzip} depicts the time required by the \bzip binary
to decompress each of the inputs generated by \pname.  We notice that for the
first hour of fuzzing, the inputs generated by \pname do not exhibit
significant slowdown during their decompression by \bzip. In
particular, each of the 250-byte inputs of \pname's corpus for the first hour of
fuzzing is decompressed in approximately 0.0006 seconds.  However, in upcoming
generations, we observe that \pname successfully achieves decompression times
reaching 0.18s to 0.21s and an overall slowdown in the range of 300x. Particularly,
in the first 6 minutes after the
first hour, \pname achieves a decompression time of 0.10 sec. This first peak in
the decompression time is achieved because of \pname triggering the
randomization mechanism of \bzip, by setting the respective header
byte to a non-zero value. This mechanism, although deprecated, was put in place
to protect against repetitive blocks, and is still supported for legacy
reasons. However, even greater slowdowns are achieved when \pname mutates two
bytes used in \bzip's Move to Front Transform
(MTF)~\cite{bzip2manual} and particularly in the run length encoding of the MTF
result. Specifically, the mutation of these bytes affects the total number of
invocations of the \texttt{BZ2\_bzDecompress} routine, which results in a total
slowdown of 38.31x in decompression time.

The respective code snippet in which the affected bytes are read is shown in
Listing~\ref{lst:bzip}: the \texttt{GET\_MTF\_VAL} macro reads the modified
bytes in memory~\footnote{Via the macros \texttt{GET\_BITS(BZ\_X\_MTF\_3, zvec,
zn)} and \texttt{GET\_BIT(BZ\_X\_MTF\_4, zj)}}. These bytes subsequently cause
the routine \texttt{BZ2\_bzDecompress} to be called 4845 times, contrary to a
single call before the mutation. We should note at this point, that the total
size of the input before and after the mutation remained unchanged.

Finally, in order to compare with a non complexity-targeting strategy, we
repeated the previous experiment using traditional coverage-based fuzzing. The
fuzzer, when guided only based on coverage, did not generate any input causing
executions larger than 0.0008 seconds, with the maximum slowdown achieved being
23.7\%.

\lstset{basicstyle=\scriptsize, style=Cstyle, gobble=-6}
    \begin{lstlisting}[caption=Excerpt from bzip2's BZ2\_decompress routine
    (decompress.c). A two byte modification by SlowFuzz results in a 38.31x
    slowdown compared to the previous input., label=lst:bzip]
 do {
   /* Check that N doesn't get too big, so that
    es doesn't go negative.  The maximum value
    that can be RUNA/RUNB encoded is equal
    to the block size (post the initial RLE),
    viz, 900k, so bounding N at 2 million
    should guard against overflow without
    rejecting any legitimate inputs. */
   if (N >= 2*1024*1024) RETURN(BZ_DATA_ERROR);
   if (nextSym == BZ_RUNA) es = es + (0+1) * N; else
   if (nextSym == BZ_RUNB) es = es + (1+1) * N;
   N = N * 2;
   GET_MTF_VAL(BZ_X_MTF_3, BZ_X_MTF_4, nextSym);
 }
   while (nextSym == BZ_RUNA || nextSym == BZ_RUNB);
\end{lstlisting}

From the above experiment we observe that \pname's guidance and mutations
engines are successful in pinpointing locations that trigger large slowdowns
even in very complex applications such as a state-of-the-art compression
utility like \bzip.

\vspace{0.1in}
\begin{mdframed}
    \textbf{Result 2}: \pname is capable of exposing complexity vulnerabilities
    (e.g., 300x slowdown in \bzip, PCRE-compliant regular expressions with
    exponential matching time, and PHP hash table collisions) in real-world,
    non-trivial applications without knowing any domain-specific details.
\end{mdframed}
\vspace{0.1in}

\subsection{Engine Evaluation}
\label{subsec:engine_eval}

\begin{figure}[ht!]
\includegraphics[width=\columnwidth]{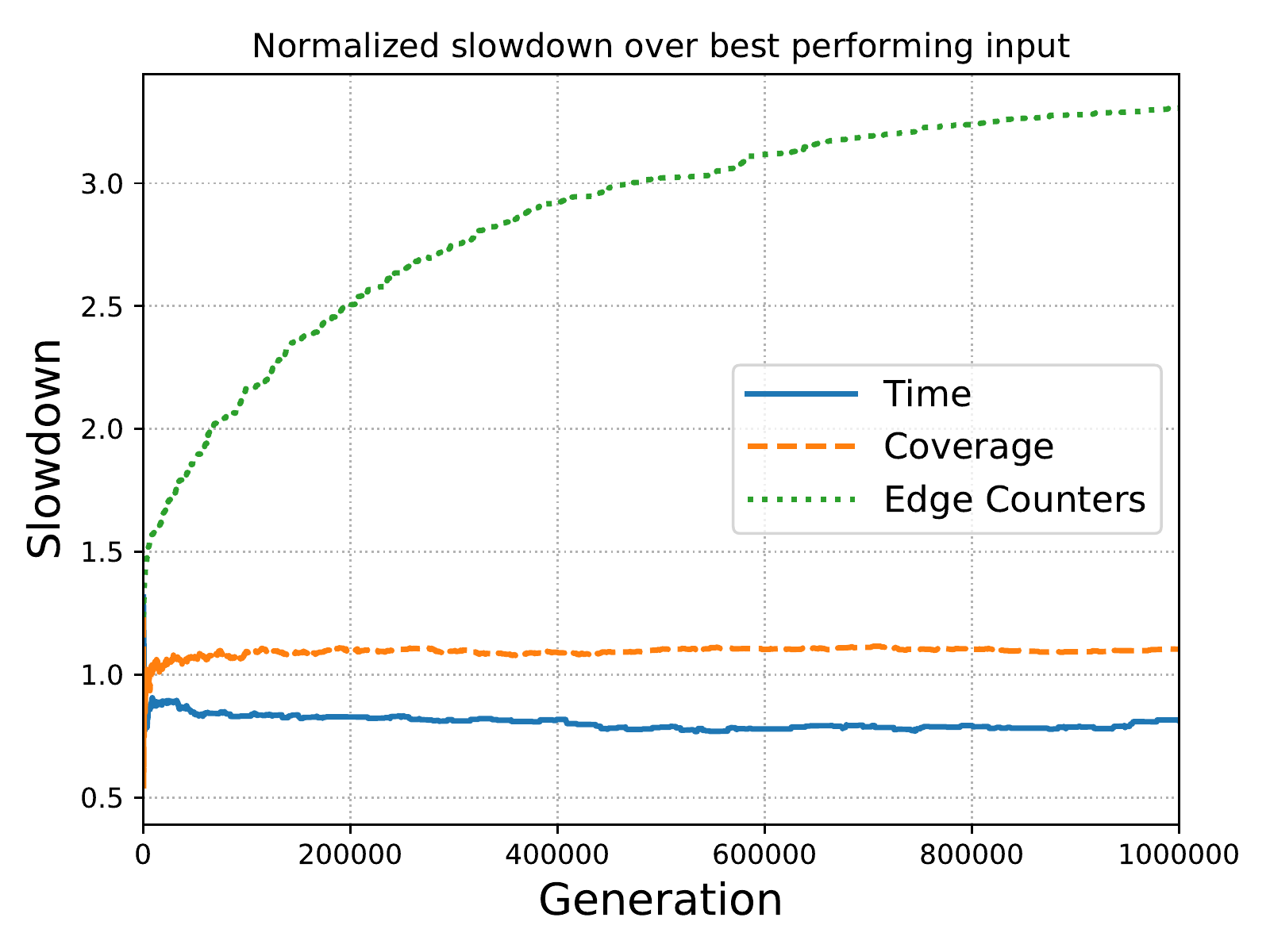}
    \caption{Comparison of the slowdown achieved by \pname under different
    guidance mechanisms, when applied on the OpenBSD quicksort
    implementation of Section~\ref{subsec:microbench}, for an input size of 64
    bytes, after 1 million generations (average of 100 runs).}
    \label{fig:engguide}
\end{figure}

\noindent
{\bf Effect of \pname's fitness function.}
\label{subsubsec:mut_engines}
In this section, we examine the effect of using code-coverage-guided search versus
\pname's resource usage based fitness function, particularly in the context of
scanning an application
for complexity vulnerabilities. To do so, we repeat one of the experiments of
Section~\ref{subsec:microbench}, applying \pname on the OpenBSD quicksort
implementation with an input size of 64 bytes, for a total of 1 million
generations, using hybrid mutations.  Our results are presented in
Figure~\ref{fig:engguide}.
We observe that \pname's guidance mechanism yields significant improvement
over code-coverage-guided search. In particular, \pname
achieves a 3.3x slowdown for OpenBSD, whereas the respective
slowdown achieved using only coverage-guided search is 23.41\%.
This is an expected result, since, as mentioned in previous Sections,
code coverage cannot encapsulate behaviors resulting in multiple invocations of
the same line of code (\eg an infinite loop). Moreover, we notice that
the total instructions of each unit that is created by \pname at different
generations is not monotonically increasing. This is an artifact of our
implementation, using SanitizerCoverage's 8-bit counters, which provide
a coarse-grained, imprecise tracking of the real number of times each edge was
invoked (Section~\ref{sec:impl}). Thus, although a unit might
result in execution of fewer instructions, it will only be observed
by \pname's guidance engine if the respective number of total CFG edges
falls into a separate bucket (8 possible ranges representing the total number
of CFG edge accesses). Future work can consider applying more precise instruction
tracking (\eg using hardware counters or utilities similar to \texttt{perf})
with static analyses passes, to achieve more effective guidance.


Finally, when choosing the \pname fitness function, we also considered the option of
utilizing time-based tracking instead of performance counters. However,
performing time-based measurements in real-world systems is not trivial,
especially at instruction-level granularity and when multiple samples are required
in order to minimize measurement errors. In the context of fuzzing,
multiple runs of the same input will slow the fuzzer down significantly.
To demonstrate this point, in Figure~\ref{fig:engguide}, we also
include an experiment in which the execution time of an input is used to guide
input generation. In particular, we utilized CPU clock time to measure
the execution time of a unit and discarded the unit if it was not slower than
all previously seen units. We notice that the corpus degrades due to system noise
and does not achieve any slowdown larger than 23\%.~\footnote{Contrary to the
slowdowns measured during fuzzing using a single run, the slowdowns presented
in Figure~\ref{fig:engguide} are generated using the \texttt{perf} utility
running ten iterations per input. Non-monotonic increases denote corpus
degradation due to bad input selection.}

\vspace{0.1in}
\begin{mdframed}
    \textbf{Result 3:} \pname's fitness function and mutation schemes outperform
    code-coverage-guided evolutionary search by more than 100\%.
\end{mdframed}
\vspace{0.1in}

\begin{figure}[ht!]
\includegraphics[width=\columnwidth]{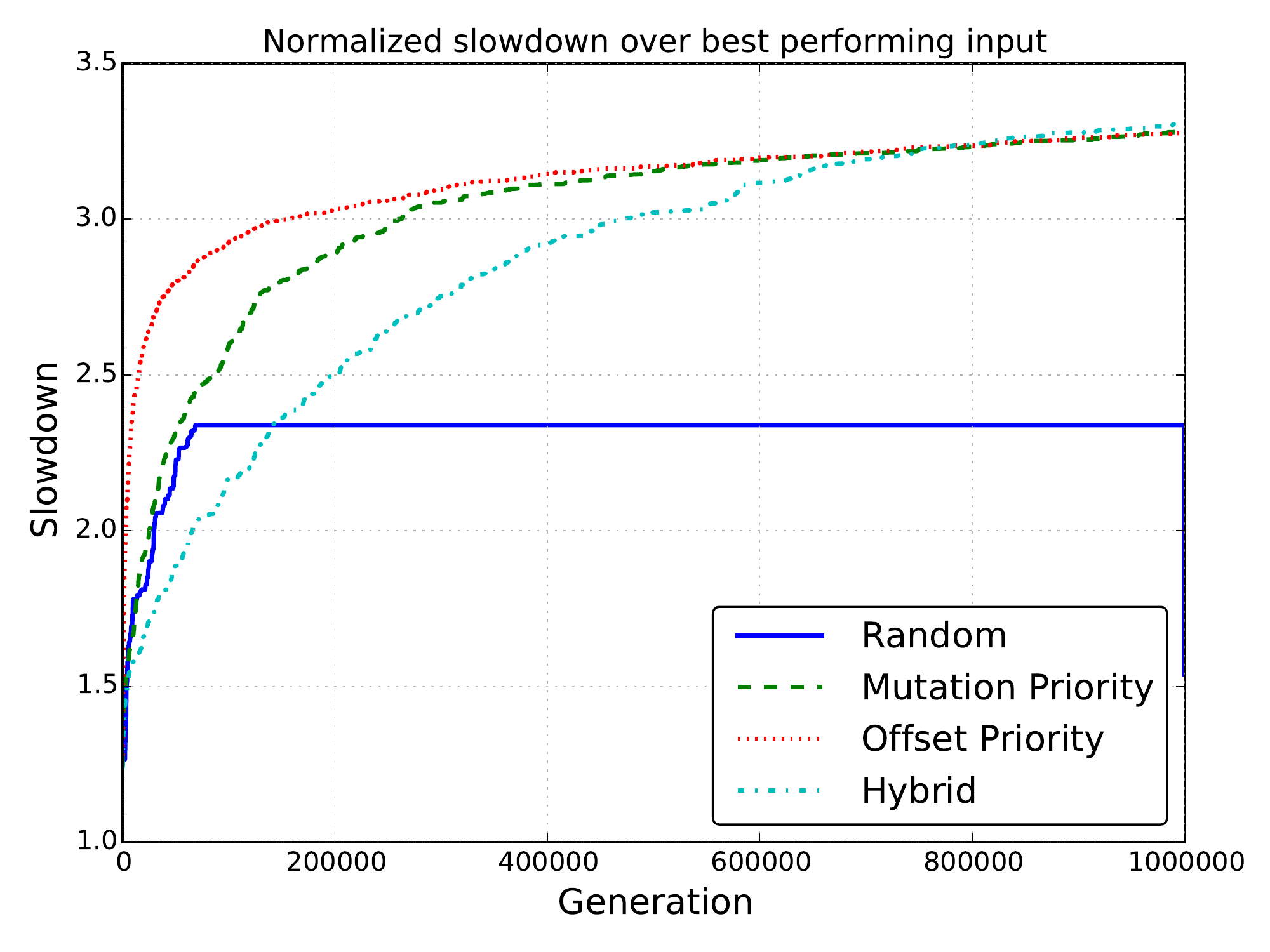}
    \caption{Comparison of the best slowdown achieved by \pname's different
    mutation schemes, at each generation, when applied on the OpenBSD
    quicksort implementation of Section~\ref{subsec:microbench}, for an input
    size of 64 bytes, after 1 million generations (average of 100 runs).}
    \label{fig:mut_engines}
\end{figure}

\noindent
{\bf Effect of Mutation Schemes.}
\label{subsubsec:mut_engines}
To highlight the different characteristics of each of \pname's mutation
schemes described in Section~\ref{sec:method}, we repeat one of the experiments of Section~\ref{subsec:microbench},
applying \pname on the OpenBSD quicksort, each time using a different mutation
strategy. Our experimental setup is identical with that of
Section~\ref{subsec:microbench}: we sort inputs with a size of 64 bytes and
fuzz for a total of 1 million generations. For each mode of operation, we
average on a total of 100 \pname sessions. Our results are presented in
Figure~\ref{fig:mut_engines}.


We notice that, for the above experiment, choosing a mutation at random, is the
worst performing option among all mutation options supported by
\pname (Section~\ref{subsec:mutations}), however still achieving a slowdown of
2.33x over the best performing input.  Indeed, all of \pname's scoring-based
mutation engines (offset-priority, mutation-priority and hybrid), are expected
to perform at least as good as selecting mutations at random, given enough
generations, as they avoid getting stuck with unproductive mutations.  We also
observe that offset priority is the fastest mode to converge out of the other
mutation schemes for this particular experiment, and results in
an overall slowdown of 3.27x.

For sorting, offsets that correspond to areas of the array that should \textit{not} be
mutated, are quickly penalized under the offset priority scheme, thus mutations are mainly performed on the non-sorted
portions of the array.  Additionally, we observe that mutation priority also
outperforms the random scheme due
to the fact that certain mutations (\eg crossover operations) may have
devastating effects on the sorting of the array. The mutation priority scheme picks
up such patterns and avoids such mutations. By contrast, these mutations continue to be used
under the random scheme.  Finally, we
observe that the hybrid mode eventually outperforms all other strategies, achieving a
3.30x slowdown, however is the last mutation mode to start reaching a plateau.
We suspect that this results from the fact the hybrid mode does not quickly penalize
particular inputs or mutations as it needs more samples of each mutation operation and offset
pair before avoiding any particular offset or mutation operation.

\noindent
{\bf Instrumentation overhead.}
\label{subsubsec:mut_engines}
\pname's runtime overhead, measured in executions per second, matches the
overhead of native libFuzzer. The executions per second achieved on different
payloads are mostly dominated by the runtimes of
the native binary, as well as the respective I/O operations. Despite our
choice to prototype \pname using libFuzzer, the design and methodology
presented in Section~\ref{sec:method} can be applied to any evolutionary
fuzzer and can also be implemented using Dynamic Binary Instrumentation
frameworks, such as Intel's PIN~\cite{pin}, to allow for more detailed runtime
tracking of the application state.  However, such frameworks are known to incur
slowdowns of more than 200\%, even with minimal instrumentation~\cite{dynaguard}.
For instance, for our PHP hashtable experiments described in Section~\ref{subsec:micro_hash},
an insertion of 16 strings,
resulting in 8 collisions, takes 0.02 seconds. Running the same insertion under
a PIN tool that only counts instructions, requires a total of
\textasciitilde 2 seconds. By contrast, hashtable fuzzing with \pname
achieves up to 4000 execs/sec, unless a
significant slowdown is incurred due to a particular
input.~\footnote{Execution under \pname does not require repeated loading of the
required libraries, but is only dominated by the function being tested, which
is only a fraction of the total execution of the native binary (thus smaller
than 0.02 seconds).}

\section{Discussion}
\label{sec:discussion}
In this paper, we demonstrated that evolutionary search techniques commonly used
in fuzzing to find memory safety bugs can be adapted to find algorithmic complexity
vulnerabilities. Similar strategies should be applicable for finding other types of
DOS attacks like battery draining, filling up memory or hard disk, etc. Designing
the fitness functions and mutation schemes for detecting such bugs will be an interesting
future research problem. Besides evolutionary techniques, using other mechanisms like
reinforcement learning or Monte Carlo search techniques can also be adapted
for finding inputs with worst-case resource usage.

Our current prototype of \pname is completely dynamic. However, integrating
static analysis techniques into \pname can further improve its performance.
Using static analysis to find potentially promising offsets in an input for mutation
will further reduce the search space and therefore will make the search process
more efficient. For example, using taint tracking and loop analysis together with
runtime flow profiles can identify potentially promising code locations that can cause
significant slowdowns~\cite{song_loopPerformance,mudduluru_flow_profiling_for_perf}.

The current prototype implementation of \pname uses the SanitizerCoverage
passes to keep track of the number of times a CFG edge is accessed.
Such tracking is limited by the total number of buckets allowed by
SanitizerCoverage. This reduces the accuracy of resource usage information
as tracked by \pname. This results from the fact that any edge that is accessed
more than 128 times is assigned to the same bucket regardless of the actual
number of accesses. Although, under its current implementation, the actual
edge count information is imprecise, this is not a fundamental design
limitation of \pname but an artifact of our prototype implementation.
Alternative implementations can offer more precise tracking can via custom
callbacks for SanitizerCoverage, by adopting hardware counters or by
utilizing per-unit \texttt{perf} tracking.
On the other hand, the benefit of the current
implementation is that it can be incorporated into libFuzzer's main engine
orthogonally, without requiring major changes to libFuzzer's dependencies.

\section{Related Work}
\label{sec:related}

\noindent
{\bf Complexity attacks.} Detecting and mitigating algorithmic complexity attacks is an
active field of research. Crosby \etal~\cite{dos_via_complexity} were the first
to present complexity attacks abusing collisions in hash table implementations.
Contrary to \pname's approach, however, their attack required expert knowledge.
Since then, several lines of
work have explored attacks and defenses targeting different types of complexity
attacks: Cai \etal~\cite{cai2009exploiting} leverage complexity vulnerabilities
in the Linux kernel name lookup hash tables to exploit race
conditions in the kernel \texttt{access(2)/open(2)} system calls, whereas Sun
\etal~\cite{sun2011covert} explore complexity vulnerabilities in the name
lookup algorithm of the Linux kernel to achieve an exploitable covert timing
channel. Smith \etal~\cite{smith2006backtracking} exploit the syntax of the
Snort IDS to perform a complexity attack resulting in slowdowns during packet
inspection. Shenoy \etal~\cite{shenoy2012hardware,shenoy2012improving}
present an algorithmic complexity attack against the popular Aho-Corasick string searching
algorithm and propose hardware and software-based defenses to mitigate the worst-case
performance of their attacks. Moreover, several lines of work focus
particularly on statically detecting complexity vulnerabilities related to
regular expression matching,
especially focusing on backtracking during the matching process~\cite{rexploiter,
berglund2014analyzing, namjoshi2010robust,kirrage2013static}. Contrary to
\pname, all the above lines of work require deep domain-dependent knowledge
and do not expand to different categories of complexity vulnerabilities.

Finally, recent work by Holland \etal~\cite{holland2016statically} combines
static and dynamic analysis to perform analyst-driven exploration of Java programs
to detect complexity vulnerabilities. However, contrary to \pname, this work
requires a human analyst to closely guide the exploration process,
specifying which portions of the binary should be analyzed statically and which
dynamically as well as defining the inputs to the binary.

\noindent
{\bf Performance bugs.} Several prior works target generic performance bugs not
necessarily related to complexity vulnerabilities. For instance, Lu \etal
study a large set of real-world performance bugs to construct a set of
rules that they use to discover new performance bugs via
hand-built checkers integrated in the LLVM compiler
infrastructure~\cite{jin_understanding_perf_bugs}. Along the same lines,
LDoctor~\cite{song_loopPerformance} detects loop inefficiencies by implementing
a hybrid static-dynamic program analysis that leverages different loop-specific
rules. Both the above lines of work,
contrary to \pname, require expert-level knowledge for creating the detection rules,
and are orthogonal to the current work.
Another line of work focuses on application profiling to detect performance bottlenecks.
For example, Ramanathan\etal
utilize flow profiling for the efficient detection of memory-related
performance bugs in Java programs ~\cite{mudduluru_flow_profiling_for_perf}.
Grechanik \etal utilize a genetic-algorithm-driven profiler for detecting
performance bottlenecks ~\cite{profilingbottleneck} in Web applications, and
cluster execution traces to explore different combinations of the input
parameter values. However, contrary to \pname, their goal is to explore a large
space of input combinations in the context of automatic application profiling
and not to detect complexity vulnerabilities.

\noindent
{\bf WCET.} Another related line of work addresses accurate
Worst-Case Execution Time (WCET) estimation for a given application.  Apart
from static analysis and evolutionary testing approaches~\cite{bernat2002wcet},
traditionally WCET estimation has been achieved using search based methods
measuring end-to-end execution times~\cite{tracey2002search}. Moreover, Hybrid
Measurement-Based Analyses (HMBA) have been used to measure the execution times
of program segments via instrumentation
points~\cite{petters1999making,petters2000bounding,betts2010hybrid} and
execution profiles~\cite{bernat2002wcet}. Wegener \etal~\cite{verifyingtiming}
utilize evolutionary techniques for testing timing constraints in real-time
systems, however contrary to \pname, apply processor-level timing measurements
for their fitness function and only perform random mutations.
Finally, recent techniques combine
hardware effects and loop bounds with genetic algorithms~\cite{khanbate}.
However, all of the above methods attempt to detect worst-case
execution times for simple and mostly straight-line program segments often used
in real-time systems. By contrast, \pname detects algorithmic complexity attacks
in large complex programs deployed in general purpose hardware.

\noindent
{\bf Evolutionary Fuzzing.} Several lines of work deploy evolutionary
mutation-based fuzzing to target crash-inducing bugs. Notable examples are the
AFL~\cite{afl}, libFuzzer~\cite{libFuzzer}, hongfuzz~\cite{hongfuzz}, and
syzkaller~\cite{syzkaller} fuzzers, as well as the CERT Basic Fuzzing Framework
(BFF)~\cite{bff}, which utilize coverage as their main guidance mechanism.
Moreover, several frameworks combine coverage-based evolutionary fuzzing with
symbolic execution ~\cite{godefroid_2008_ndss_sage,
stephens_2016_ndss_driller, dowsing,
sangkilcha_2015_snp_mutationalfuzzing}, or with static analysis and
dynamic tainting~\cite{vuzzer} to achieve higher code coverage and increase
their effectiveness in detecting bugs. Finally, NEZHA~\cite{nezha} utilizes
evolutionary-based, mutation-assisted testing to target semantic bugs.
Although many of the aforementioned lines of research share many common
building blocks with \pname, they do not target complexity vulnerabilities and
mainly utilize random mutations contrary to \pname's targeted mutation
strategies.

\section{Conclusion}
\label{sec:concl}

In this work we designed \pname, the first, to the best of our knowledge, evolutionary-search-based
framework targeting algorithmic complexity vulnerabilities. We evaluated \pname on
a variety of real-world applications including zip utilities, regular expression libraries and hash table
implementations. We demonstrated that \pname can successfully generate inputs
that match the theoretical worst-case complexity in known algorithms. We also showed
that \pname was successful in triggering complexity vulnerabilities in all the
applications we examined. \pname's evolutionary engine and
mutation strategies generated inputs causing more than 300-times slowdown in
the \bzip decompression routine, produced inputs triggering high numbers of
collisions in production-level hash table implementations, and also generated
regular expressions with exponential matching complexities without any
knowledge about the semantics of regular expressions. We believe
our results demonstrate that customized evolutionary search techniques present
a promising direction for automated detection of not only algorithmic
complexity vulnerabilities, but also of other types of resource exhaustion
vulnerabilities, and hope to aspire tighter integration of existing techniques
and static analyses with modern mutation-based evolutionary testing.

\section{Acknowledgments}
We would like to thank the anonymous reviewers for their valuable feedback.
This work is sponsored in part by the Office of Naval Research (ONR) grant
N00014-17-1-2010, the National Science Foundation (NSF) grants CNS-13-18415 and
CNS-16-17670, and a Google Faculty Fellowship. Any opinions, findings,
conclusions, or recommendations expressed herein are those of the authors, and
do not necessarily reflect those of the US Government, ONR, NSF, or Google.
\bibliographystyle{acm}
\bibliography{paper}
\appendix





\section{WAF Regexes}
\label{sec:waf_regexes}

The slowdowns presented in Figure~\ref{fig:all_wafs} correspond to inputs matched
against the following regular expressions:

{\bf Regex 1:}
\lstset{basicstyle=\scriptsize, style=plain, numbers=none}
\begin{lstlisting}
(?i:(j|(&#x?0*((74)|(4A)|(106)|(6A));?)) 
([\t]|(&((#x?0*(9|(13)|(10)|A|D);?)| 
(tab;)|(newline;))))*(a|(&#x?0*((65)| 
(41)|(97)|(61));?))([\t]|(&((#x?0*(9| 
(13)|(10)|A|D);?)|(tab;)|(newline;)) 
))*(v|(&#x?0*((86)|(56)|(118)|(76));?) 
)([\t]|(&((#x?0*(9|(13)|(10)|A|D);?)| 
(tab;)|(newline;))))*(a|(&#x?0*((65)| 
(41)|(97)|(61));?))([\t]|(&((#x?0*(9| 
(13)|(10)|A|D);?)|(tab;)|(newline;))))* 
(s|(&#x?0*((83)|(53)|(115)|(73));?))( 
[\t]|(&((#x?0*(9|(13)|(10)|A|D);?)| 
(tab;)|(newline;))))*(c|(&#x?0*((67)| 
(43)|(99)|(63));?))([\t]|(&((#x?0*(9| 
(13)|(10)|A|D);?)|(tab;)|(newline;))))* 
(r|(&#x?0*((82)|(52)|(114)|(72));?)) 
([\t]|(&((#x?0*(9|(13)|(10)|A|D);?)| 
(tab;)|(newline;))))*(i|(&#x?0*((73)| 
(49)|(105)|(69));?))([\t]|(&((#x?0*(9| 
(13)|(10)|A|D);?)|(tab;)|(newline;))))* 
(p|(&#x?0*((80)|(50)|(112)|(70));?)) 
([\t]|(&((#x?0*(9|(13)|(10)|A|D);?)| 
(tab;)|(newline;))))*(t|(&#x?0*((84)| 
(54)|(116)|(74));?))([\t]|(&((#x?0*(9| 
(13)|(10)|A|D);?)|(tab;)|(newline;)))) 
*(:|(&((#x?0*((58)|(3A));?)|(colon;) 
))).)
\end{lstlisting}
{\bf Regex 2:}
\begin{lstlisting}
<(a|abbr|acronym|address|applet|area|
audioscope|b|base|basefront|bdo| 
bgsound|big|blackface|blink| 
blockquote|body|bq|br|button|caption| 
center|cite|code|col|colgroup| 
comment|dd|del|dfn|dir|div|dl| 
dt|em|embed|fieldset|fn|font| 
form|frame|frameset|h1|head|hr| 
html|i|iframe|ilayer|img|input|ins|  
isindex|kdb|keygen|label|layer| 
legend|li|limittext|link|listing| 
map|marquee|menu|meta|multicol| 
nobr|noembed|noframes|noscript| 
nosmartquotes|object|ol|optgroup| 
option|p|param|plaintext|pre|q| 
rt|ruby|s|samp|script|select| 
server|shadow|sidebar|small| 
spacer|span|strike|strong|style| 
sub|sup|table|tbody|td|textarea| 
tfoot|th|thead|title|tr|tt|u|ul| 
var|wbr|xml|xmp)\\W
\end{lstlisting}
{\bf Regex 3:}
\begin{lstlisting}
(?i:<.*[:]vmlframe.*?[ /+\t]*?src[ /+\t]*=)
\end{lstlisting}

\end{document}